\documentclass[aps,pre,epsf,superscriptaddress,amsmath,amssymb,amsfonts,twocolumn,showpacs]{revtex4-1}

\usepackage[utf8]{inputenc}

\usepackage{amsmath}
\usepackage{amssymb}
\usepackage{amsfonts}
\usepackage{amstext}
\usepackage{amsthm}
\usepackage{mathtools}
\usepackage{physics}

\usepackage{bm} 

\usepackage{graphicx}

\usepackage{epsfig}
\usepackage{dcolumn}
\usepackage{bm}
\usepackage{braket}
\usepackage{color}
\usepackage[colorlinks=true,citecolor=blue,urlcolor=blue]{hyperref}

\usepackage[english]{babel}

\usepackage{multirow}

\usepackage{xcolor}

\definecolor{deeppurple}{rgb}{0.7, 0, 0.8}

\allowdisplaybreaks
\advance\parskip 0.4pt

\usepackage{natbib}

\begin{document}

\title{On-demand generation of dark-bright soliton trains in Bose-Einstein condensates}
\author{A. Romero-Ros}
\affiliation{Center for Optical Quantum Technologies, Department of Physics,
University of Hamburg, Luruper Chaussee 149, 22761 Hamburg,
Germany}

\author{G. C. Katsimiga}
\affiliation{Center for Optical Quantum Technologies, Department of Physics,
University of Hamburg, Luruper Chaussee 149, 22761 Hamburg,
Germany}
\affiliation{The Hamburg Centre for Ultrafast Imaging,
University of Hamburg, Luruper Chaussee 149, 22761 Hamburg,	Germany}

\author{P. G. Kevrekidis}
\affiliation{Department of Mathematics and Statistics, University
of Massachusetts Amherst, Amherst, MA 01003-4515, USA}

\author{B. Prinari}
\affiliation{Department of Mathematics, State University
	of New York, Buffalo, New York 14260, USA}

\author{G. Biondini}
\affiliation{Department of Mathematics, State University of New York, Buffalo, New York 14260, USA}
\affiliation{Department of Physics, State University of New York, Buffalo, New York 14260, USA}

\author{P. Schmelcher}
\affiliation{Center for Optical Quantum Technologies, Department of Physics,
University of Hamburg, Luruper Chaussee 149, 22761 Hamburg, Germany}
\affiliation{The Hamburg Centre for Ultrafast Imaging,
University of Hamburg, Luruper Chaussee 149, 22761 Hamburg, Germany}

\date{\today}

\begin{abstract}
   The controlled creation of dark-bright (DB) soliton trains in
   multi-component Bose-Einstein condensates (BECs) is a topic of
   ongoing interest.
   In this work we generalize earlier findings on the creation of dark soliton trains in single-component BECs \href{https://journals.aps.org/pra/abstract/10.1103/PhysRevA.103.023329}{[A. Romero-Ros \textit{et al}., Phys. Rev. A \textbf{103}, 023329 (2021)]} to two-component BECs.
   By choosing suitable \textit{filled} box-type initial configurations (FBTCs) and solving the direct scattering problem for the defocusing vector nonlinear Schrödinger equation with nonzero boundary conditions we obtain analytical expressions for the DB soliton solutions produced by a general FBTC.
   It is found that the size of the initial box and the amount of filling directly affects the number, size, and velocity of the solitons, while the initial phase determines the parity (even or odd) of the solutions.
   Our analytical results are compared to direct numerical integration of the coupled Gross-Pitaevskii equations, both in the absence and in the presence of a harmonic trap, and an excellent agreement between the two is demonstrated.
\end{abstract}

\maketitle

\section{Introduction} \label{sec:intro}

Nonlinear phenomena in Bose-Einstein condensates (BECs) have become a focus of attention during the last couple of decades~\cite{Pethick2008,Pitaevskii2016}, and solitary waves stemming from the balance between dispersion and the nonlinearities of the system have been a topic of intense investigation~\cite{Kevrekidis2015}.
In single-component BECs, these macroscopic nonlinear excitations can have the form of local density suppressions (dark solitons~\cite{Frantzeskakis2010,Kevrekidis2015}) or local density humps (bright solitons~\cite{Abdullaev2005}) depending on whether the nonlinear interaction is repulsive or attractive, respectively.

The experimental realization of two-component BECs~\cite{Myatt1997,Hall1998,Stamper-Kurn1998} has opened a  window towards the study of more complex solitonic structures~\cite{Busch2001,Ohberg2001,Kevrekidis2004,Dabrowska-Wuster2007,Hoefer2011,Danaila2016,Qu2016,Kevrekidis2016,Morera2018,Katsimiga2020}.
In repulsive two-component BECs, a fundamental excitation takes the
form of a dark-bright (DB) soliton~\cite{Busch2001}.
A single DB soliton consists of a dark soliton that acts as an effective potential in which the bright soliton is trapped and, consequently, waveguided.
Importantly, bright solitons cannot be sustained (unless under such
waveguiding) in self-repulsive BECs.
The concept of waveguiding has its origin in nonlinear
optics~\cite{Sheppard1997,Ostrovskaya1998} (see also references
therein), where DB solitons have  been an active topic of theoretical
and experimental research~\cite{Kivshar1998,Ostrovskaya1999,Kivshar2003}.
In this context, the DB soliton dynamics is described by the
defocusing vector nonlinear Schrödinger (VNLS)
equation~\cite{Kevrekidis2015}, while in the context of BECs DB
solitons similarly obey the so-called coupled Gross-Pitaevskii equation (CGPE)~\cite{Gross1961,Pitaevskii1961,Pethick2008}.

The first experimental realizations of DB solitons in BECs
almost a decade
ago~\cite{Becker2008,Middelkamp2011,Yan2011,Hoefer2011,Hamner2011,Hamner2013},
as well as subsequent experimental realizations of their
variants and generalizations~\cite{Danaila2016,Bersano2018,Katsimiga2020,Farolfi2020,Chai2021,Lannig2020},
have motivated a significant amount of interest 
in studying their
dynamics and
interactions~\cite{Rajendran2009,Yin2011,Achilleos2011,Alvarez2011,Alvarez2013,Wang2015,Biondini2015b,Katsimiga2017,Katsimiga2017a,Morera2018,Katsimiga2018,Alotaibi2018,Alotaibi2019,Arazo2021}.
In particular, several methods have been proposed to create DB soliton structures.
For instance, the combination of phase imprinting techniques~\cite{Burger1999,Denschlag2000}, to create the dark soliton, and a local population transfer by means of a Raman process~\cite{Dum1998}, to create the bright counterpart, allows the creation of individual DB solitons~\cite{Becker2008}.
Other population transfer methods demonstrated how an alternating spatial distribution of the two components, via the creation of a winding pattern, can lead to the formation of DB soliton trains ~\cite{Hamner2013}.
Additionally, counterflow techniques which involve a dynamical mixing
of both components also give rise to DB soliton
trains~\cite{Hamner2011,Hoefer2011}.
More recently, the controllable creation of DB pairs could generate
the
conditions for a systematic observation and measurement of their
interactions, including in BECs with a higher (e.g., three) number of
components~\cite{Lannig2020}.

Following the counterflow concept, matter-wave interference methods have been highly used in single-component BECs to generate dark soliton trains~\cite{Weller2008,Hoefer2009,Reinhardt1997, Scott1998,Theocharis2010}.
This method is based on the collision of two separated condensates, and allows for the systematic nucleation of a  desired number of solitonic entities upon tailoring the initial separation of the colliding condensates and their relative phase.
In this counterflow setting, exact results were originally derived for the defocusing NLS equation in the seminal work of Ref.~\cite{Zakharov1973} for a box-type pulse by means of the inverse scattering transform (IST)~\cite{Faddeev2007,Espinola-Rocha2009,Demontis2013,Biondini2014,Biondini2015a}.
More recently, some theoretical works have exploited the integrable nature of the defocusing VNLS model and further developed an IST formalism with non-zero-boundary conditions (NZBC)~\cite{Prinari2006,Prinari2011,Biondini2015b}.

In view of our previous work in single-component
BECs~\cite{Romero-Ros2021} and the analytical tools provided by the
direct scattering method and the IST with NZBC, in this work we
exploit the unprecedented level of control that the ultracold
environment offers~\cite{Myatt1997,Hall1998,Inouye1998,Stamper-Kurn1998,Chin2010}
to study the response of a two-component, one-dimensional (1D),
harmonically trapped BEC with repulsive intracomponent and intercomponent
interactions, when a general \textit{filled} box-type configuration (FBTC) is considered as an initial condition.
In particular, in our setup the wave function of the first component is a box-type pulse whose sides play the role of the two colliding condensates in the matter-wave interference mechanism.
On the other hand, the wave function of the second component is an inverted box that \textit{fills} the space between the two sides of the box of the first component (see Fig.~\ref{fig:IC}).
A somewhat similar configuration (albeit with differences
in the bright component) was  considered in nonlinear
optics to study vector soliton interaction dynamics~\cite{Ostrovskaya1999}.
First, we consider the integrable version of the problem, i.e, the defocusing VNLS equation with NZBC.
Here, we solve analytically the direct scattering problem for the aforementioned box-type configuration and provide the discrete eigenvalues of the scattering problem for distinct parametric variations.
The latter characterize the amplitudes and velocities of the ensuing DB solitons, whose exact waveform can be then extracted via the IST.

Having at hand the exact analytical expressions for the DB solitons,
we then compare them with direct numerical simulations of the CGPE
with a FBTC in the absence of confinement, finding remarkable
agreement, as should be expected on the basis of the exact nature of
the IST analysis.
Moreover, to showcase the broader, as well as physical relevance of
our results, we extend our analytical findings to the case involving the
presence of a harmonic confinement.
Using the expressions for the eigenvalues from the direct scattering problem, we design analytical estimates to describe the in-trap oscillation dynamics of the generated DB solitons.
Here, we provide explicit expressions accounting for the oscillating size of the dark and bright counterparts of a DB soliton in a trap.
The latter is a feature that is absent in the single-component case, which we attribute to the intercomponent interaction.
An excellent agreement between the analytical estimates and the
numerical simulations confirms the extension of the predicted
solutions of the direct scattering problem from the homogeneous setup
to the harmonically trapped scenario. This also justifies the
particular relevance and usefulness of the detailed IST analysis of
the integrable case with a view towards the more physically relevant confined setting.

Our presentation is organized as follows.
In Sec.~\ref{sec:model} we introduce the model and solve the direct scattering problem for the defocusing VNLS equation with a general FBTC.
Additionally, we discuss some analytical considerations regarding the eigenvalues of the scattering problem and the DB soliton solution.
In Sec.~\ref{sec:solutions} we present our findings.
First, we extract the eigenvalues of the scattering problem for a wide range of different initial configurations.
Then, we perform a direct comparison between our analytical findings and the numerical integration of the CGPE, both in the absence and in the presence of a harmonic trap.
Finally, in Sec.~\ref{sec:conclusions} we summarize our results and discuss possible directions for future study.
In Appendix~\ref{app:DB_solutions} we provide further details on the DB soliton solutions.
In Appendix~\ref{app:amplitudes} we describe the change of amplitude of oscillating DB solitons in the presence of a trap.

\section{Nonlinear Schrödinger Equation and Dark-Bright Soliton Solution} \label{sec:model}

We consider a one-dimensional (1D) pseudo-spinor BEC consisting of two different spin states, e.g., $\ket{F,m}=\ket{1,-1}$ and $\ket{F,m}=\ket{2,2}$, of the same atomic species of $^{87}$Rb \cite{Myatt1997}, confined in a highly anisotropic trap with longitudinal and transverse trapping frequencies satisfying the relation $\omega_x\ll\omega_\perp$.
In such a cigar-shaped geometry, the condensate wavefunction along the transverse direction, being the ground state of the respective harmonic oscillator, can be integrated out.
This, in turn, leads to the following pair of coupled Gross-Pitaevskii equations (CGPEs) \cite{Kevrekidis2015}:
\begin{equation}
    i\hbar\partial_t\Psi_j={\mathcal{H}}_0\Psi_j+\sum_{k=1}^2 g_{jk}^{(1\textrm{D})}|\Psi_k|^2\Psi_j  \,, 
    \label{eq:CGPE}
\end{equation}
with $j=1,2$, which, in the mean-field framework, governs the BEC dynamics for the longitudinal part of the wavefunction.
In the above expression, ${\mathcal{H}}_0=-\frac{\hbar^2}{2m}\partial^2_x+V(x)$ is the single-particle Hamiltonian, where $m$ denotes the atomic mass and $V(x)=m\omega_x^2 x^2/2$ denotes the external harmonic potential.
Also, $g_{jk}^{(1\textrm{D})}=2 a_{jk}\hbar^2/ma_\perp^2$ accounts for the effective one-dimensional repulsive interaction strengths, with $a_{jk}>0$ denoting  the 1D scattering length and $a_\perp=\sqrt{\hbar/m\omega_\perp}$ being the transverse harmonic oscillator length.
Under the following transformations, 
$\Tilde{t}=t\omega_\perp$, 
$\Tilde{x}=x a_\perp^{-1}$, and
$\Tilde{q}_j=\Psi_j \sqrt{2a_\perp}$,
Eq.~\eqref{eq:CGPE} can be rewritten in the dimensionless form
\begin{equation}
    i\partial_t q_j=\qty[-\frac{1}{2}\partial^2_x+\frac{1}{2}\Omega^2x^2] q_j+\sum_{k=1}^2g_{jk}^{(1\textrm{D})}|q_k|^2 q_j \,.
    \label{eq:CGPE_adim}
\end{equation}
Here, $\Omega\equiv\omega_x/\omega_{\perp}$ and $g_{jk}^{(1\textrm{D})}=a_{jk}/a_\perp$.
Note that for convenience we dropped the tildes and that energy, time and length are now measured in units of $\hbar\omega_\perp$, $\omega_\perp^{-1}$ and $a_\perp=\sqrt{\hbar/m\omega_\perp}$, respectively.

In this work, we consider $g_{jk}^{(1\textrm{D})}=1$, i.e., we work with the
classical Manakov model~\cite{Manakov1973} in the case of the absence of confinement.
Then, Eq.~\eqref{eq:CGPE_adim}, with $\Omega=0$, reduces to the vector nonlinear Schrödinger (VNLS) equation, namely, 
\begin{align}
    i\vb{q}_t+\frac{1}{2}\vb{q}_{xx}-\norm{\vb{q}}^2\vb{q}=0 \,,
    \label{eq:NLS}
\end{align}
to which we can further perform the rescaling $\Tilde{\vb{q}}(x,t)=\vb{q}(\sqrt{2}x,t)\exp{-2i q_o^2t}$ that leads, by dropping the tilde, to
\begin{align}
    i\vb{q}_t+\vb{q}_{xx}-2(\norm{\vb{q}}^2-q_o^2)\vb{q}=0 \,,
    \label{eq:manakov_system}
\end{align}
which is subject to the following time-independent NZBC at infinity
\begin{align}
    \lim_{x\to\pm\infty} \vb{q}(x,t)=\vb{q_\pm}=\vb{q}_oe^{i\theta_\pm} \,.
\end{align}
Hereafter, $\vb{q} \equiv \vb{q}(x,t)$ and $\vb{q}_o$ are
two-component vectors, $\norm{\cdot}$ is the standard Euclidean norm,
$q_o=\norm{\vb{q}_o}>0$, $\theta_\pm$ are real numbers, and subscripts
$x$ and $t$ denote partial differentiation with respect to space and time  hereafter.

\begin{figure}[t]
    \centering
    \includegraphics[width=\columnwidth]{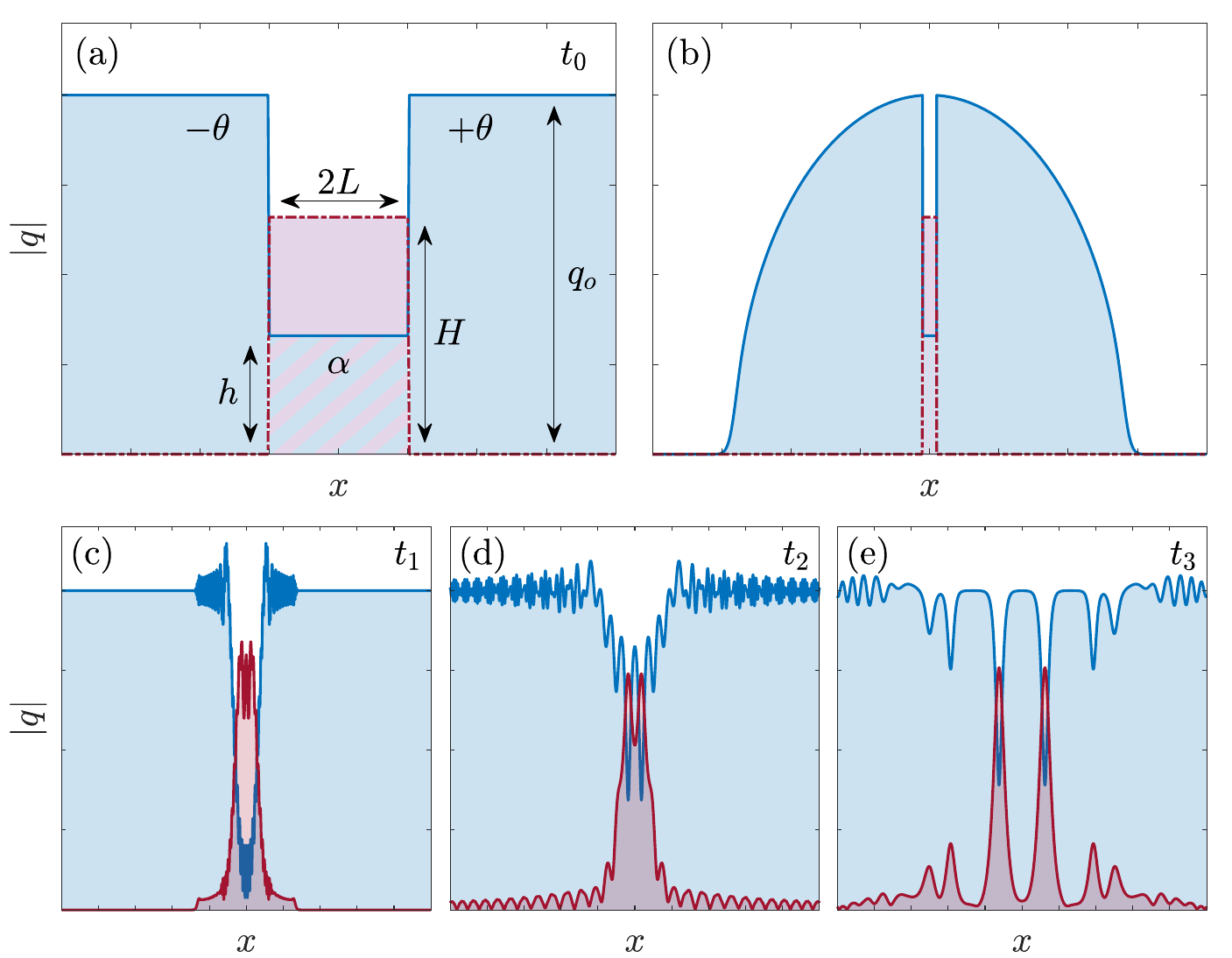}
    \caption{Schematic illustration of the box-type configuration utilized herein, for arbitrary $L,\,q_o,\, \theta,\, h$, $\alpha$ and $H$ in the absence (a) and in (b) the presence of a harmonic trapping potential.
   	Here, $L$ is the parameter that controls the separation between the two sides of the box which play the role of the two colliding condensates in a matter-wave interference process.
   	The matter-wave interference process is also schematically illustrated through snapshots in (c), (d) and (e) at times $t_1<t_2<t_3$, respectively. 
    Note that the quantities shown are measured in transverse oscillator units.
}
    \label{fig:IC}
\end{figure}

{Building on our recent investigation of scalar
  BECs~\cite{Romero-Ros2021}, here we consider a box-type initial
  configuration in the first component whose \textit{box} is being
  filled by the second component (so that the latter can induce the
  formation
  of bright solitons) in the following manner}
\begin{align}
    \vb{q}(x,0) =
\left\{
	\begin{array}{ll}
		(q_oe^{-i\theta},0)^T\,,   & \quad x < -L  \\
		(he^{i\alpha},H)^T\,,     & \quad \abs{x} < L \\
		(q_oe^{i\theta},0)^T\,,   & \quad x > L 
	\end{array}
\right.
\label{eq:initial_conditions}
\end{align}
A schematic illustration of Eq.~\eqref{eq:initial_conditions} is given in Fig.~\ref{fig:IC}(a).
Here, $0\leq h\leq q_o$ refers to the height of the box of the first component, and $0\leq H\leq q_o$ refers to the height of the filling box of the second component.
$q_o$ is the amplitude of the box, $\theta_\pm$ are the phases on each side of the box and $\alpha$ is the phase of the first component inside the box.
The phase invariance of the VNLS equation allowed us to define $\theta_+=-\theta_-=\theta$ without loss of generality in Eq.~(\ref{eq:initial_conditions}).
For convenience, we further introduce the quantities
\begin{equation}
\Delta\theta = 2\theta\,,\qquad
\Delta\theta_- = \theta+\alpha\,,\qquad
\Delta\theta_+ = \theta-\alpha\,, 
\label{eq:phase_jump}
\end{equation}
to denote the distinct phase differences in each of the different regions of the box.
We will refer to the cases $\Delta\theta=0$ and $\Delta\theta=\pi$ as in-phase (IP) and out-of-phase (OP) configurations, respectively, and to the special case having $h=0$ as the ``zero-box" configuration, which describes the absence of atoms of the first component inside the box.

Additionally, $L$ corresponds to the half width of the box and it is the parameter that controls the distance between the two sides of the box playing the role of the two-colliding condensates in the matter-wave interference mechanism.
A schematic illustration of the latter is shown through snapshots in Figs.~\ref{fig:IC}(c)--(e) at $t_1<t_2<t_3$, respectively.
At $t_1$ the two sides of the box are spreading towards each other and form an interference pattern inside the box.
Then, at $t_2$, some of the fringes formed due to the interference process stabilize and start acting as effective potentials for the second component filling the box.
Finally, at $t_3$ the stabilized fringes develop into dark solitons, while the second component trapped inside the latter becomes bright solitons, giving rise to a DB soliton train.  

\subsection{The direct scattering problem} \label{sec:scattering_problem}

The defocusing VNLS equation [see Eq.~\eqref{eq:manakov_system}], corresponds to a coupled system of integrable nonlinear partial differential equations that can be solved analytically by means of the IST in terms of a Lax pair.
The $3\times 3$ Lax pair associated with Eq.~\eqref{eq:manakov_system} is
\begin{align}
    \bm{\phi}_x=\vb{X}\bm{\phi}\,, \qquad \bm{\phi}_t=\vb{T}\bm{\phi}\,,
    \label{eq:scattering _problem}
\end{align}
where $\bm{\phi}$ is a $3\times 3$ matrix eigenvector,
\begin{align}
    \vb{X}(x,t,k)&=ik\vb{J}+\vb{Q}\,,
    \\
    \vb{T}(x,t,k)&=2ik^2\vb{J}-i\vb{J}(\vb{Q}_x-\vb{Q}+q_o^2)-2k\vb{Q}\,,
    \label{eq:X_T}
\end{align}
with
\begin{align}
    \vb{J}=\mqty(-1 & \vb{0}^T \\ \vb{0} & \vb{I})\,, \qquad 
    \vb{Q}(x,t)=\mqty(0 & \vb{q}^T \\ \vb{q}^* & \vb{0}) \,,   
    \label{eq:J_Q}
\end{align}
and $\vb{I}$ and $\vb{0}$ are the appropriately sized identity and zero matrix, respectively.
The first equation in Eq.~\eqref{eq:scattering _problem} is referred to as the scattering problem and $k \in \mathbb{C}$ as the scattering parameter.

Under fairly general conditions on $\vb{q}(x,t)$, as $x\to\pm\infty$ the solutions of the direct scattering problem are approximated by those of the asymptotic scattering problems $\bm{\phi}_x=\vb{X}_\pm\bm{\phi}$, where $\vb{X}_\pm=ik\vb{J}+\vb{Q}_\pm$ and $\vb{Q}_\pm=\lim_{x\to\pm\infty}\vb{Q}(x,t)$.
The eigenvalues of $\vb{X}_{\pm}$ are $ik$ and $\pm i\lambda$, where
\begin{align}
    \lambda(k) = \sqrt{k^2-q_o^2} \,.
\end{align}
These eigenvalues have branch points, and therefore we introduce the two-sheeted Riemann surface defined by $\lambda(k)$.
As in Refs.~\cite{Prinari2006,Biondini2014,Biondini2015,Biondini2015a}, we take the branch cut along the semilines $(-\infty,-q_o)$ and $(q_o,\infty)$, and we label those sheets such that $\Im\lambda(k)\geq0$ on sheet I and  $\Im\lambda(k)\leq0$ on sheet II.

We also define the Jost solutions, $\bm{\phi}_\pm(x,t,k)$, as the simultaneous solutions of both parts of the Lax pair satisfying the boundary conditions
\begin{align}
    \bm{\phi}_\pm(x,t,k)\equiv \vb{Y}_\pm(k)e^{i\vb{\Theta}(x,t,k)} + {\mathcal{O}}(1) 
    \label{eq:jost_solution}
\end{align}
as $x\to\pm\infty$, where $\vb{\Theta}(x,t,k)=\vb{\Lambda}x-\vb{\Omega}t$ with $\vb{\Lambda}=\text{diag}(-\lambda,k,\lambda)$, $\vb{\Omega}=\text{diag}(2k\lambda, -(k^2+\lambda^2), -2k\lambda)$, and
$\vb{Y}_\pm(k)$ are the simultaneous eigenvector matrices of $\vb{X}_\pm$ and $\vb{T}_\pm=\lim_{x\to\pm\infty}\vb{T}(x,t,k)$.
The two sets of Jost solutions are related to each other through the scattering relation
\begin{align}
    \bm{\phi}_-(x,t,k) = \bm{\phi}_+(x,t,k) \vb{S}(k) \,,
\end{align}
valid for all $k \in (-\infty,-q_o)\bigcup(q_o,\infty)$.
Moreover, the fact that $\bm{\phi}_\pm$ are simultaneous solutions of both parts of the Lax pair implies that the scattering coefficients and the discrete eigenvalues of the scattering operator are time-independent.
Therefore, hereafter we will consider the scattering problem at $t=0$ and we will omit the time dependence from the eigenfunctions.

At $t=0$ the scattering problem in each of the three regions $x<-L$, $|x|<L$, and $x>L$ takes the form $\bm{\phi}_x=(ik\vb{J}+\vb{Q}_j)\bm{\phi}$ with the index $j=c,\pm$ and constant potentials $\vb{Q}_\pm$ and $\vb{Q}_c$ given by
\begin{subequations}
\label{eq:Q}
    \begin{gather}
        \vb{Q}_\pm = \mqty( 0 & q_oe^{\pm i\theta} & 0 \\
                        q_oe^{\mp i\theta} & 0 & 0 \\
                        0 & 0 & 0 ) \,,
        \\
        \vb{Q}_c = \mqty( 0 & he^{i\alpha} & H \\
                        he^{-i\alpha} & 0 & 0 \\
                        H & 0 & 0 ) \,.
    \end{gather}
\end{subequations}
One can then easily find explicit solutions for the scattering problem in each of the aforementioned regions, namely,
\begin{subequations}
\begin{align}
    \bm{\varphi}_l(x,k) &= \vb{Y}_-(k)e^{i\vb{\Lambda} x} \qquad x\le -L \\
    \bm{\varphi}_c(x,k) &= \vb{Y}_c(k)e^{i\vb{M} x} \qquad |x|\le L \\
    \bm{\varphi}_r(x,k) &= \vb{Y}_+(k)e^{i\vb{\Lambda} x} \qquad x\ge L
\end{align}
\label{eq:regions}
\end{subequations}
where $\vb{M}=\text{diag}(-\mu,k,\mu)$, $\mu=\sqrt{k^2-(h^2+H^2)}$ and 
\begin{subequations}
    \begin{gather}
        \vb{Y}_{\pm}(k) = \mqty( \lambda+k & 0 & \lambda-k \\
                                 iq_oe^{\mp i\theta} & 0 & -iq_oe^{\mp i\theta} \\
                                 0 & iq_oe^{\pm i\theta} & 0) \,,
        \\
        \vb{Y}_c(k) = \mqty( \mu+k & 0 & \mu-k \\
                                 ihe^{-i\alpha} & -iH & -ihe^{-i\alpha} \\
                                 iH & ihe^{i\alpha} & -iH) \,.
        \label{eq:Y}
    \end{gather}
\end{subequations}
Equations \eqref{eq:regions} yield explicit representations for the Jost solutions $\bm{\phi}_\pm(x,0,k)$ in their respective regions, i.e., $\bm{\phi}_{-}(x,0,k)\equiv \bm{\varphi}_l(x,k)$ for $x \leq -L$, and 
$\bm{\phi}_{+}(x,0,k)\equiv \bm{\varphi}_r(x,k)$ for $x\geq L$.
At the boundary of each region one can express the fundamental solution on the left as a linear combination of the fundamental solution on the right, and vice versa.
In particular, we can introduce scattering matrices $\vb{S}_{-}(k)$ and $\vb{S}_{+}(k)$ such that
\begin{subequations}
    \label{eq:parcial_scattering_matrices}
    \begin{align}
        \bm{\varphi}_- (-L,k) &= \bm{\varphi}_c(-L,k) \vb{S}_- (k) \,, \\
        \bm{\varphi}_c (L,k) &= \bm{\varphi}_+(L,k) \vb{S}_+ (k) \,.
    \end{align}
\end{subequations}
As a consequence, we can express the scattering matrix $\vb{S}(k)$ relating the Jost solutions $\bm{\phi}_\pm(x, k)$ as
\begin{align}
    \vb{S}(k) &= \vb{S}_+(k)\vb{S}_-(k) \nonumber \\
              &= e^{-i\vb{\Lambda} L} \vb{Y}_+^{-1}\vb{Y}_c
                 e^{2i\vb{M}L}\vb{Y}_c^{-1}\vb{Y}_-e^{-i\vb{\Lambda} L} \,.
        \label{eq:scatterting_matrix}
\end{align}
Computing the right-hand side of Eq.~\eqref{eq:scatterting_matrix}, we obtain the following expression for the first element, $s_{11}(k)$, of the scattering matrix $\vb{S}(k)$
\begin{align}
    \label{eq:s11}
    4\lambda\mu & q_o(h^2+H^2)e^{-2i\lambda L}s_{11}(k)=  \nonumber \\
    & =4ih(h^2+H^2)q_o^2e^{i\theta}\cos\alpha\sin(2\mu L) \nonumber \\
    & +2q_oh^2e^{2i\theta}(\lambda-k)[\mu\cos(2\mu L)+ik\sin(2\mu L)] \nonumber \\
    & + 2q_o(h^2+H^2)(\lambda+k)[\mu\cos(2\mu L)-ik\sin(2\mu L)] \nonumber \\
    & +2q_o\mu H^2(\lambda-k)e^{2i\theta}e^{2ikL} \,.
\end{align}
The discrete eigenvalues of the scattering problem are the zeros of $s_{11}(k)$ for all $k\in\mathbb{C}$ with $\Im \lambda(k)>0$, where $s_{11}(k)$ is analytic~\cite{Prinari2006}.
It is important to remark that, in general, for the defocusing VNLS equation the eigenvalues of the scattering problem are not only single zeros, but double zeros can also occur \cite{Biondini2015a}.
However, for the particular configuration used in this work [see Eq.~\eqref{eq:initial_conditions}] all zeros will turn out to be simple.

\subsection{{The Dark-Bright soliton solution}}

In view of the inverse problem, it is convenient to introduce a uniformization variable $z$ defined by
\begin{align}
    z=k+\lambda \,,
    \label{eq:z}
\end{align}
which is inverted by
\begin{align}
    k=\frac{1}{2}\qty(z+\frac{q_o^2}{z})\,, \qquad \lambda = \frac{1}{2}\qty(z-\frac{q_o^2}{z})\,.
    \label{eq:inverted_z}
\end{align}
Thereby, sheets I and II of the Riemann surface are mapped onto the upper and lower half-planes of the complex $z$ plane, respectively;
the continuous spectrum is [i.e., the semilines $(-\infty,-q_0)\cap
(q_0,\infty)$ are]  mapped onto the real $z$ axis, while the spectral gaps $(-q_0,q_0)$ on both sheets are mapped onto the circle of radius $q_0$ (see Ref.~\cite{Prinari2006} for further details).
The discrete eigenvalues are found as zeros of
$s_{11}(z):=s_{11}(k(z),\lambda(z))$, and in this case a zero of
$s_{11}(z)$ on the upper semicircle of radius $q_o$ corresponds to a
dark-dark soliton, i.e, a dark soliton in each component, while a zero
inside the upper semicircle of radius $q_o$ corresponds to a DB
soliton.
In the presence of a single such zero, the inverse scattering problem yields the following DB soliton solution \cite{Prinari2006}:
\begin{widetext}
    \begin{subequations}
        \label{eq:DB_solution}
        \begin{align}
            q_d(x,t)&= \Big\{q_o\cos{\beta_o} -iq_o\sin{\beta_o}\tanh{\big[\nu_o(x-x_0+2\xi_o t)\big]}\Big\}e^{i(\beta_o +\varphi_d +2q_o^2t)} \,,
            \label{eq:dark_solution}
            \\
            q_b(x,t)&= -i\sin{\beta_o} \sqrt{q_o^2-\abs{z_o}^2}\sech{\big[\nu_o(x-x_0+2\xi_o t)\big]}e^{i(\xi_o x -(\xi_o^2-\nu_o^2)t+\varphi_b + 2q_o^2t)} \,, 
            \label{eq:bright_solution}
        \end{align}
    \end{subequations}
\end{widetext}
as a solution of Eq.~\eqref{eq:manakov_system}.
Here, $q_d$ is the dark soliton component and $q_b$ is the bright one.
Also, $x_0$ is the center of the soliton and $\varphi_{d,b}$ are arbitrary constant phases.
The DB solution of Eq.~\eqref{eq:DB_solution} is expressed in terms of the spectral parameter $z_o=\abs{z_o}e^{i\beta_o}\equiv \xi_o+i\nu_o$, with
\begin{align}
    \xi_o = \abs{z_o}\cos{\beta_o} \,, \qquad \nu_o=\abs{z_o}\sin{\beta_o} \,.
    \label{eq:xi_nu}
\end{align}
Therefore, the relevant soliton parameters can be uniquely specified in terms of $z_o$, i.e.,
\begin{subequations}
    \label{eq:DB_parameters} 
    \begin{align}
        A_d = q_o\sin\beta_o & \equiv \frac{q_o}{|z_o|}\Im z_o \,, 
        \label{eq:Ad}
        \\
        A_b = \sqrt{q_o^2-|z_o|^2}\sin{\beta_o} & \equiv \Im z_o\sqrt{\frac{q_o^2}{|z_o|^2}-1} \,, \label{eq:Ab}
        \\
        v = -2|z_o|\cos \beta_o & \equiv -2\Re z_o \,, 
        \label{eq:v}
    \end{align} 
\end{subequations}
where $A_d$ and $A_b$ are the dark and bright soliton amplitudes, respectively, and $v$ denotes the DB soliton velocity.

Equivalently, the soliton parameters can be directly expressed in terms of $k_o$ (see Appendix~\ref{app:DB_solutions}).
Given a zero $k_o$, one can substitute $z_o=k_o+\lambda_o$ into Eqs.~\eqref{eq:DB_parameters} with the caveat that $\lambda_o=\sqrt{k_o^2-q_o^2}$ {must be chosen with the appropriate branch cut, and on the appropriate branch} where $\Im \lambda_o >0$. 
Then, Eqs.~\eqref{eq:DB_parameters} become
\begin{subequations}
    \label{eq:DB_parameters_gamma}
    \begin{gather}
        A_d = \frac{2\gamma \Im \lambda_o}{\gamma^2+1} \,, \\
        A_b = -\frac{2\Im k_o}{\sqrt{\gamma^2-1}} \,, \\
        v = -\frac{4\Re k_o}{1+\gamma^2} \,,
    \end{gather}
    where
    \begin{align}
        \gamma=\frac{q_o}{|z_o|}>1 \,. \nonumber
    \end{align}
\end{subequations}

\begin{figure}[t]
    \centering
    \includegraphics[width=\columnwidth]{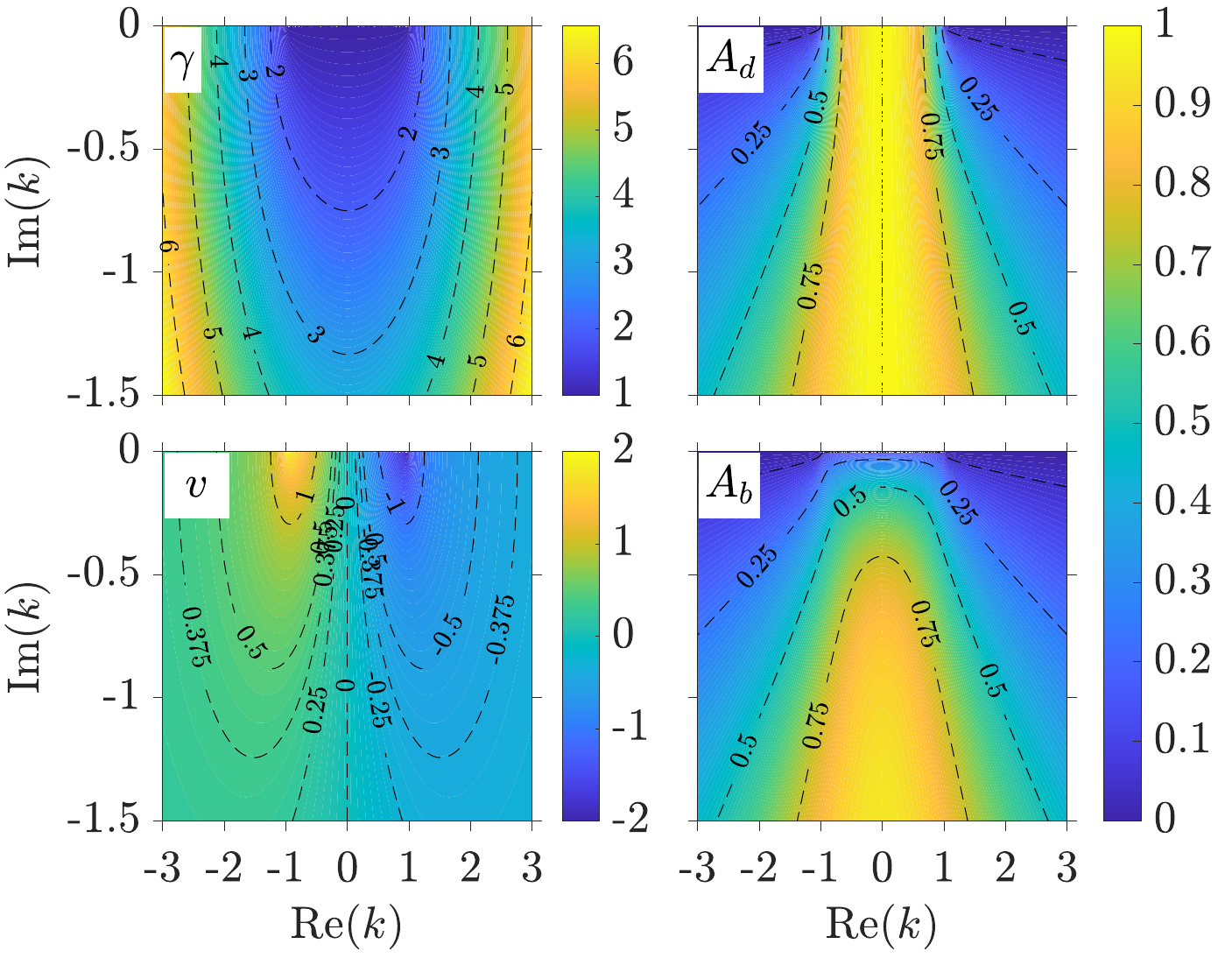}
    \caption{$\gamma$ and the soliton parameters amplitudes, $A_d$, $A_b$, and velocity, $v$, as functions of the scattering parameter $k$ for $q_o=1$.
    Note that the quantities shown are measured in transverse oscillator units.}
    \label{fig:DB_param_q}
\end{figure}

To get some physical insight on the DB solutions, we illustrate in Fig.~\ref{fig:DB_param_q} the dependence parameters $\gamma$, $A_d$, $A_b$ and $v$ on the scattering parameter $k$, for the solutions provided by Eq.~\eqref{eq:gamma_solutions_real} and for $q_o=1$.
Here, one can see that indeed $\gamma>1 \ \forall \ k$.
Also, $A_d \leq q_o \ \forall \ k$, as expected, since dark solitons cannot have amplitudes greater than the background.
Similarly,  $A_b < A_d \ \forall \ k$. 
Obviously, larger (deeper) dark solitons can host larger bright solitons, but in turn the DB soliton itself becomes slower.
In fact, $v$ has a minimum ($v=0$) at $\Re k_o = 0$, where $A_d$ has a maximum ($A_d=q_o$).
The latter is known as a black soliton, and it can host a bright
soliton of any smaller size, which explains why $v$ and $A_d$ are independent of $\Im k$ at $\Re k = 0$.
On the other hand, $v$ always has its maximum (absolute) value at $k=2q_o$, coinciding with the speed of sound of the condensate, $c=2q_o$ note that $c=\sqrt{gn}$ \cite{Bogoliubov1947,Lee1957}, where $n$ is the peak density of the BEC, in the dimensionless units adopted herein for the CGPE~\eqref{eq:CGPE_adim}].
Yet, no soliton solution exists with $v=c$.
Further details on the soliton parameters are discussed in Sec.~\ref{sec:analytic_solutions}.

%

\section{Dark-Bright soliton generation and dynamics} \label{sec:solutions}

\subsection{Analytical results for the discrete spectrum} \label{sec:analytic_solutions}

In this section, we aim at finding the zeros of $s_{11}(k)$ [see Eq.~\eqref{eq:s11}] and analytically characterizing the DB solitons produced by the FBTC in Eq.~\eqref{eq:initial_conditions}, upon considering different variations of the system parameters.
In particular, our initial FBTC is defined by six different parameters: the half width, $L$, the amplitude, $q_o$, the side phases, $\pm\theta$, the height, $h$, of the first component in the box, its phase, $\alpha$, and the filling of the second component in the box, $H$.
The corresponding values of our parameter exploration are the following:
\begin{align}
    & L\in[1,9] \,,\quad   \theta = \qty{0,\frac{\pi}{2}}\,, \nonumber \\
    & h\in[0,q_o]\,,\quad  \alpha = \qty{0,\pi} \,,\quad H\in[0,q_o] \,, \nonumber
\end{align}
together with $q_o=1$. 
Furthermore, { we introduce the filling angle, $\sigma \in [0,\pi]$, which relates the heights $h$ and $H$ with the amplitude background $q_o$ as follows:}
\begin{subequations}
        \label{eq:sigma}
    \begin{align}
        h = q_o\cos\sigma \,,
        \\
        H = q_o\sin\sigma \,.
    \end{align}
\end{subequations}
Introducing $\sigma$ allows us to explore different filling configurations using a single parameter.
Notice that $h>0$ in the regime $0\leq \sigma < \pi/2$, while $h<0$ in the regime $\pi/2 < \sigma \leq \pi$, which is equivalent to $h=|q_o\cos\sigma|>0$ with $\alpha = \pi$ (see Fig.~\ref{fig:IC}).

Since we are mostly interested in effects driven by the presence of the second component, we choose $\sigma$ as our main parameter.
We are also interested in considering the effect of distinct initial configurations, and we take $L$ as our second representative parameter since it controls the separation between the colliding sides of the condensate.
Thus, below we will vary $\sigma$ for different values of $L$, denoting such variation as $L[\sigma]$.

To classify all spectra, we choose two different case scenarios.
The first one consists of a zero-box configuration ($h=0$), where the second component is the only component present inside the box.
The second case is a full-box configuration, with the box being fully filled either by a single or by both components, i.e., $q_o^2 = h^2(\sigma)+H^2(\sigma)$ [see Eqs.~\eqref{eq:sigma}].
We start exploring IP-FBTCs ($\theta=0$), followed by OP-FBTCs ($\theta=\pi/2$).

The corresponding spectra of zeros are presented in Figs.~\ref{fig:roots_LH_q1_th0_h0_a0}, \ref{fig:roots_LH_q1_thpi2_h0_a0}, \ref{fig:roots_LhH_th_0}, and \ref{fig:roots_LhH_th_pi2}.
All these figures share the same arrangement.
In particular, each figure consists of 10 panels, (a)--(j), distributed along two rows and five columns.
The latter correspond to five different values of $L$, ranging from $L=1$ to 9.
The top row shows the zeros of $s_{11}(k)$ in the $\Re k - \sigma$ plane with $\Im k$ depicted as a color gradient in a logarithmic scale.
This representation provides a clearer disposition of the zeros.
Additionally, zeros corresponding to $\alpha=0$ ($\alpha=\pi$) are shown on a white (gray) background [see, e.g., Fig.~\ref{fig:roots_LH_q1_th0_h0_a0} (Fig.~\ref{fig:roots_LhH_th_0})]. 
In contrast, bottom-row panels depict the zeros in the complex $k$-plane, which can be directly mapped onto Fig.~\ref{fig:DB_param_q}, containing the relevant physical information of the solitons, such as their amplitudes and velocities.
In this case, $\sigma$ is illustrated as a color gradient.
Blue tones ($0\leq\sigma\leq\pi/2$) correspond to $\alpha=0$, while red tones ($\pi/2 \leq\sigma\leq\pi$) correspond to $\alpha=\pi$.
Recall that, in all cases, $\sigma$ is the main varying parameter and both rows can be easily compared by following their common $\Re k$ axis.
Also, when looking at the zeros, e.g. in Fig.~\ref{fig:roots_LH_q1_th0_h0_a0}, one should keep in mind that in this system under consideration the zeros are single-valued, i.e., $k_i\neq k_j$, where $i,j$ denote different zeros, for any choice of parameters (see Sec.~\ref{sec:scattering_problem}).
This means that although some of the zeros appear to be on top of each other they never intersect, i.e. coincide, which is the case since our two-dimensional representation of the zeros, e.g., in Fig.~\ref{fig:roots_LH_q1_th0_h0_a0}, is a projection of a three-dimensional space ($\Re k$, $\Im k$, $\sigma$).
Finally, if both spatial and phase symmetries of the FBTC are preserved, the zeros appear in pairs $k_\pm$, i.e., $\Re k_{+}=-\Re k_-$ and $\Im k_{+} = \Im k_-$.
Note that the FBTC is always spatially symmetric [see
Eq.~\eqref{eq:initial_conditions}], and thus only FBTCs with $\theta
\neq 0$ and $h \neq 0$ can present asymmetric solutions
(Fig.~\ref{fig:roots_LhH_th_pi2}).
In those cases, we say that the phase symmetry
of the system is broken.

Therefore, whenever $\theta = 0$ (Figs.~\ref{fig:roots_LH_q1_th0_h0_a0} and \ref{fig:roots_LhH_th_0}) or both symmetries are preserved (Fig.~\ref{fig:roots_LH_q1_thpi2_h0_a0}), only $k_o$ with $\Re k_o \geq 0$ are shown.

\subsubsection{\textbf{Zero-box configuration}} \label{sec:zero_box_analytic}

For the zero-box configuration we set $h = 0$,  so that only the second component is present inside the box of the FBTC.
At the same time, for our particular choice of parameters, the FBTC preserves both spatial and phase symmetries, independently of $\theta$, and thus also do its solutions.
In particular, IP-FBTCs ($\theta=0$) always present an even  number of paired zeros ($k_\pm$).
On the other hand, for this zero-box configuration, OP-FBTCs ($\theta=\pi/2$) always possess a particular zero, $k_0\in\mathbb{I}$, which is unpaired, resulting in an odd number of zeros.
More specifically, as we later explain, $k_0$ corresponds to a static DB soliton, with the dark counterpart being a so-called black soliton ($v=0$ and $A_d=q_o$).
Also note the distinct subscript 0 used when compared to \textit{o} introduced for a general solution.

\begin{figure*}[t]
    \centering
    \includegraphics[width=\linewidth]{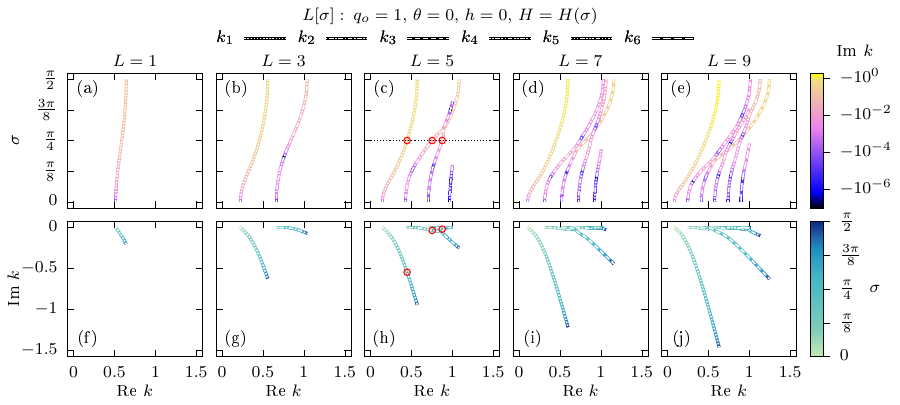}
    \caption{Zeros of $s_{11}(k)$ as a function of $\sigma$ for different values of $L$ in the zero-box IP background configuration.
    The parameters $q_o=1$, $\theta=0$, $h=0$ ($\alpha=0$) remain fixed.
    The upper row shows the location of the zeros in the $\Re k$-$\sigma$ plane whereas the bottom row shows the location of the zeros in the complex $k$-plane.
    The complex $k$-plane can be mapped onto Fig.~\ref{fig:DB_param_q} to retrieve the relevant physical information about the soliton solutions.
    The color coding shows the corresponding complementary quantity $\Im k$ (upper row) and $\sigma$ (bottom row).
    Only $\Re k_o>0$ are shown due to the symmetry of the zeros.
    Red circles in (c) and (h) correspond to the zeros shown in Fig.~\ref{fig:GPE_IP}.
    Note that the quantities shown are measured in transverse oscillator units.}
    \label{fig:roots_LH_q1_th0_h0_a0}
\end{figure*}
\begin{figure}[t]
    \centering
    \includegraphics[width=\columnwidth]{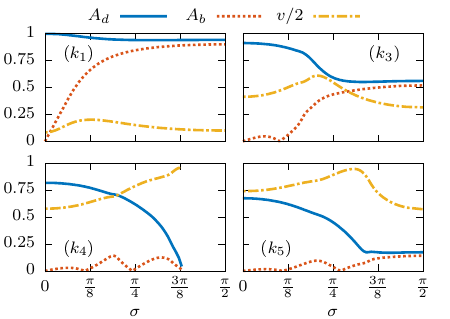}
    \caption{Amplitudes $A_d$, $A_b$, and velocity $v$ of $k_1$, $k_3$, $k_4$, and $k_5$ shown in Figs.~\ref{fig:roots_LH_q1_th0_h0_a0}(e) and \ref{fig:roots_LH_q1_th0_h0_a0}(j) (see legend) as a function of $\sigma$.
    The local maximum of $v$ defines the transition point from LIC to HIC solutions for $k_1$, $k_3$, and $k_5$.
    Note that $v$ is halved to depict all parameters in the same scale.
    Note also that the quantities shown are measured in transverse oscillator units.
    }
    \label{fig:adabv}
\end{figure}
\begin{figure*}[t]
    \centering
    \includegraphics[width=\linewidth]{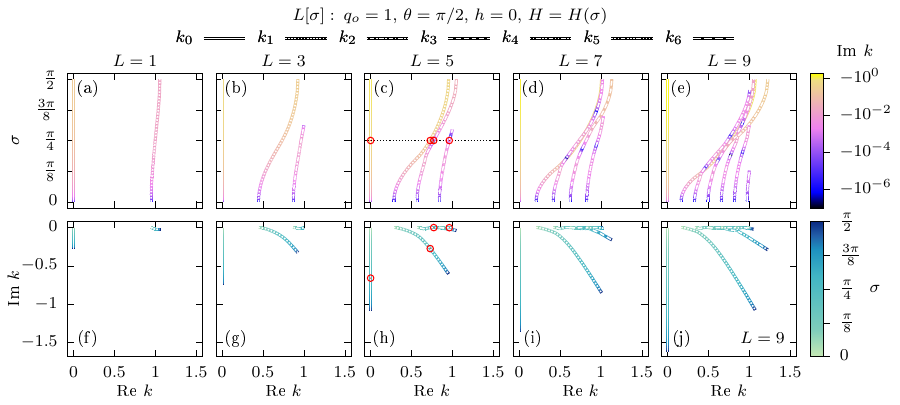}
    \caption{Zeros of $s_{11}(k)$ as a function of $\sigma$ for different values of $L$ in the zero-box OP background configuration.
    The parameters $q_o=1$, $\theta=0$, $h=0$ ($\alpha=0$) remain fixed.
    The upper row shows the location of the zeros in the $\Re k-\sigma$ plane whereas the bottom row shows the location of the zeros in the complex $k$-plane.
    The complex $k$-plane can be mapped onto Fig.~\ref{fig:DB_param_q} to retrieve the relevant physical information about the soliton solutions.
    The color coding shows the corresponding complementary quantity $\Im k$ (upper row) and $\sigma$ (bottom row).
    Only $\Re k_o>0$ are shown due to the symmetry of the zeros.
    $k_0$ is an unpaired solution.
    Red circles in (c) and (h) correspond to the zeros shown in Fig.~\ref{fig:GPE_OP}.    
    Note that the quantities shown are measured in transverse oscillator units.}
    \label{fig:roots_LH_q1_thpi2_h0_a0}
\end{figure*}

\paragraph{\textbf{In-phase background.}}

We begin by exploring the spectra of an IP zero-box configuration, for which $q_o=1$, $\theta=\pi/2$ and $h=0$ ($\alpha=0$) are held fixed.
Additionally, $L\in[1,9]$ and $\sigma\in[0,\pi/2]$, and thus $H(\sigma)\in[0,q_o]$.
The corresponding spectra of zeros are presented in Fig.~\ref{fig:roots_LH_q1_th0_h0_a0}.

From Figs.~\ref{fig:roots_LH_q1_th0_h0_a0}(a)--(e) (top row), it can be directly inferred that increasing $L$ increases the number of zeros, and thus the number of solitons, 
an outcome analogous to the single-component case \cite{Romero-Ros2021}.
In particular, $L=1$ has only one pair of zeros, $k_1$, while $L=5$
has up to four pairs, $k_1,\dots,k_4$, and $L=9$ has up to six pairs, $k_1,\dots,k_6$.
On the other hand, increasing $\sigma$ (or equivalently increasing
$H$) reduces the number of zeros.
For example, in Fig.~\ref{fig:roots_LH_q1_th0_h0_a0}(c) the spectrum of solutions goes from four pairs of zeros at $\sigma=0$ ($H=0$) to two at $\sigma=\pi/2$ ($H=1$).
Here, $k_4$ ceases to exist right above $\sigma=\pi/8$, and $k_3$ is absent for $\sigma>3\pi/8$.
We attribute this effect to an increase of the second component in the box, hindering the emergence of solitonic structures due to the repulsive intercomponent interaction.
It is also important to understand how these parametric variations affect the characteristics of the solitonic entities, in particular their amplitudes and velocities [see Eq.~\eqref{eq:DB_parameters_gamma}].
In this regard, Figs.~\ref{fig:roots_LH_q1_th0_h0_a0}(f)--(j) (bottom row) are key towards easily mapping the zeros onto the relevant physical parameters of the solitons, shown in Fig.~\ref{fig:DB_param_q}.
Although in the complex $k$-plane most of the zeros with a low imaginary contribution fall on top of each other (without intersecting), we can still use Figs.~\ref{fig:roots_LH_q1_th0_h0_a0}(a)--(e) to follow, respectively, the zeros in Figs.~\ref{fig:roots_LH_q1_th0_h0_a0}(f)--(j) by means of their common $\Re k$ axis.
It is clear that $\Re k_o$ increases with $\sigma$.
However, to infer about the behavior of $\Im k_o$ it is convenient to distinguish between solutions with a high imaginary contribution (HIC) and those with a low imaginary contribution (LIC).
We empirically define as HIC the solutions whose zeros have $|\Im k_o| > 0.1$, and LIC the ones having $|\Im k_o| < 0.01$.
First, let us focus on the LIC solutions.
In Figs.~\ref{fig:roots_LH_q1_th0_h0_a0}(f)--(j), LIC solutions lay on $\Im k_o \approx 0$, indiscernible from one another.
All solutions belong to this group when $\sigma\approx 0$ since $\Im k_o$ is a quantity directly related to the presence of the second component.
In particular, LIC solutions correspond to DB solitons with a negligible bright contribution (see Fig.~\ref{fig:DB_param_q}), i.e, they are almost pure dark solitons (see also Fig.~\ref{fig:GPE_OP} below).
However, as $\sigma$ increases, $v$ increases but $A_d$ decreases.
Similarly, $\Re k_o$ also increases with $\sigma$ and some LIC solutions cease to exist right before reaching $\Re k_o = 1$ and $\Im k_o = 0$, or equivalently, before the solitons acquire the speed of sound.
In order to avoid this point, a LIC solution must transition into a HIC one.
In Fig.~\ref{fig:roots_LH_q1_th0_h0_a0}(c), the former zeros are $k_3$ and $k_4$, and the latter are $k_1$ and $k_2$.
Additionally, in the zero-box configuration, LIC solutions present localized \textit{drops} of the imaginary contribution as $\sigma$ increases.
At these drops, the imaginary contribution drastically decreases to $\Im k_o\approx0$ to rapidly increase again.
The drops are depicted by the logarithmic colorscale of $\Im k_o$, where $\Im k_o$ drops from $|\Im k_o| \lesssim 10^{-3}$ (violet) to $|\Im k_o| \lesssim 10^{-6}$ (blue), given our numerical precision
[e.g., for $L=3$, an increase in the numerical precision leads to $\Im k_2(\sigma=1.4839988)=1.98648\times10^{-18}$].
For example, in Fig.~\ref{fig:roots_LH_q1_th0_h0_a0}(e), $k_2$ presents three drops (the region where the drops take place is blue).
The first drop takes place at $\sigma \approx 0.17\frac{\pi}{2}$.
Note that as $\sigma$ increases, drops of $k_3$, $k_4$ and $k_5$ follow.
The second drop of $k_2$ appears at $\sigma \approx 0.45\frac{\pi}{2}$.
Again, drops of $k_4$ and $k_5$ follow.
Notice that in this case, $k_3$ has already transitioned into a HIC solution (yellow tones).
The last drop takes place at $\sigma \approx 0.72\frac{\pi}{2}$.
In this case, neither $k_4$ nor $k_5$ present a drop since the former ceases to exist shortly after and the latter transitions into a HIC solution.
Nevertheless, these drops do not represent any major additional change to the solitonic structures, since $A_d$ is almost independent of $\Im k_o$ when $|\Re k_o| < 1$ and $\Im k_o\approx 0$ (see Fig.~\ref{fig:DB_param_q}), and $A_b$ is almost negligible.
A visualization of the above discussion is presented in Fig.~\ref{fig:adabv} for the LIC solution $k_4$.
Here, the effect of the drops is clearly visible on $A_b$, which decreases (almost) to zero at each drop.
Additionally here, one can appreciate how $k_4$ becomes sonic, i.e., $v\approx c$, at $\sigma\approx3\pi/8$ with a fast decrease of $A_d$ towards $A_d=0$, characteristic of the LIC solutions.

Next we focus on HIC solutions and take again as representative examples Figs.~\ref{fig:roots_LH_q1_th0_h0_a0}(e) and \ref{fig:roots_LH_q1_th0_h0_a0}(j).
Here, the HIC solutions are $k_1$, $k_3$ and $k_5$, which become more evident after they transition from LIC to HIC solutions as $\sigma$ increases.
Mapping the zeros of Fig.~\ref{fig:roots_LH_q1_th0_h0_a0}(j) onto Fig.~\ref{fig:DB_param_q} reveals that HIC solutions are DB solitons with a higher bright contribution than LIC ones. Recall that the bright contribution increases with $\Im k$.
As stated before for LIC solutions, when $\sigma$ increases $A_d$ decreases and $v$ increases, while $A_b$ increases or decreases depending on the increase or decrease of the imaginary contribution.
However, the behaviour of HIC solutions is different.
Indeed, by following $k_1$, which is a HIC solution already from low values of $\sigma$, it is obvious that this zero highly differs from the LIC solutions presented before.
In particular, it quickly reaches a regime where the ratio $\Re k_o/\Im k_o$ is almost constant independently of $\sigma$.
This regime is where we start to consider a zero as a HIC solution and, when mapped onto Fig.~\ref{fig:DB_param_q}, we observe that $k_1$ has an almost constant $A_d$.
On the other hand, in this regime $A_b$ always increases while $v$ always decreases.
The latter directly shows that DB solitons with the same dark component but a bigger bright counterpart are slower than those with a smaller bright contribution.
In the case of $k_3$, it is found that this solution transitions from a LIC to a HIC one for $\sigma > \pi/8$.
From this point onwards, the ratio $\Re k_o/\Im k_o$ also becomes almost constant, and so does again the $A_d$ related to it.
In this case, $A_b$ increases and $v$ decreases as well.
The same holds true for $k_5$, which transitions from a LIC to a HIC solution around $\sigma=3\pi/8$.

Also in this case, Fig.~\ref{fig:adabv} offers a visualization of the HIC solutions, i.e., $k_1$, $k_3$ and $k_5$.
Here, it can be seen that, once all LIC solutions have transitioned into HIC ones, they present an $A_d$ plateau.
On the other hand, it is also clear that the increase of $A_b$ directly affects $v$, which starts decreasing right before $A_d$ reaches its constant value.
Therefore, it is possible to define the transition point from LIC to HIC solution not only by the saturation of $A_d$, but also from the local maximum of $v$.

\paragraph{\textbf{Out-of-phase background.}}

We next explore the OP zero-box configuration, again for $q_o=1$, $L\in[1,9]$, $h=0$ ($\alpha=0$), and $\sigma\in[0,\pi/2]$, corresponding to $H(\sigma)\in[0,q_o]$.
However, we now fix $\theta=\pi/2$, namely setting the sides of the box out-of-phase ($\Delta\theta = \pi$).
The corresponding spectrum of zeros is illustrated in Fig.~\ref{fig:roots_LH_q1_thpi2_h0_a0}.
Again, the choice of parameters presents a symmetric distribution of the zeros and thus only $\Re k_o > 0$ are shown.

In the zero-box configuration, the peculiarity of an OP-FBTC with $\theta=\pi/2$ is that it gives rise to an odd number of solutions due to the presence of an unpaired static DB soliton, labeled $k_0$.
Note that $k_0$ is a HIC solution with $\Re k_0=0$.
Therefore, it is straightforward to map its velocity and amplitudes.
Indeed, from Fig.~\ref{fig:DB_param_q} one obtains that $v=0$ and $A_d=1$ independently of the value of $\sigma$.
On the other hand, $A_b$ increases with $\sigma$.
Aside from this extra unpaired solution, OP-FBTCs have an additional difference when compared to the IP-FBTC case.
With the OP-FBTC, as $L$ increases, the emergence of the paired zeros,
$k_1, \dots, k_6$, is slightly delayed (parametrically)
when compared to the IP-FBTC case.
This means that for some values of $L$, there are less paired zeros in the OP case than in the IP case.
For example, for $L=5$ [Fig.~\ref{fig:roots_LH_q1_thpi2_h0_a0}(c)] there exist three paired zeros, i.e., $k_1$, $k_2$, and $k_3$, contrary to the IP case [Fig.~\ref{fig:roots_LH_q1_th0_h0_a0}(c)] where also a fourth paired solution, i.e., $k_4$, was identified.

Lastly, before proceeding to the full-box configuration, it is worth
commenting  on how the presence of a second component in the box affects the solutions when compared to the single-component case.
As discussed before, when $\sigma=0$ ($H=0$) the single-component case is retrieved and the zeros identified herein coincide with the ones found in Ref.~\cite{Romero-Ros2021}, for both the IP and the OP cases.
However, as $\sigma$ increases ($H$ increases) and the box is filled with the second component, the interaction between the components prevents the emergence of all the single-component solutions, an effect which is more enhanced for overlapping components, as we will see in what follows.

\subsubsection{\textbf{Full-box configuration}} \label{sec:full_box_analytic}

For the full-box configuration we use Eq.~\eqref{eq:sigma}.
This implies that the box of the FBTC is always fully filled, either with one or both components, $q_o^2=h^2(\sigma)+H^2(\sigma)$.
By doing so, we are able to explore several configurations and elucidate the effect of the second component inside the box.
In this regard, it is important to distinguish the regimes where $H>|h|$ or $H<|h|$ and $h>0$ or $h<0$.
Besides, for $\sigma=\pi/2$ ($H=q_o$ and $h=0$) we recover the zeros from the zero-box configuration.
As before, below we explore both IP-FBTC and OP-FBTC using the previously introduced notation and labeling.
Recall that in the former case $\theta=0$ and thus the symmetry of the system is preserved, leading to a symmetric set of solutions.
On the other hand, $\theta=\pi/2$ breaks the phase symmetry of the
system, leading in turn to asymmetric solutions.

\paragraph{\textbf{In-phase background.}}

\begin{figure*}[t]
    \centering
    \includegraphics[width=\linewidth]{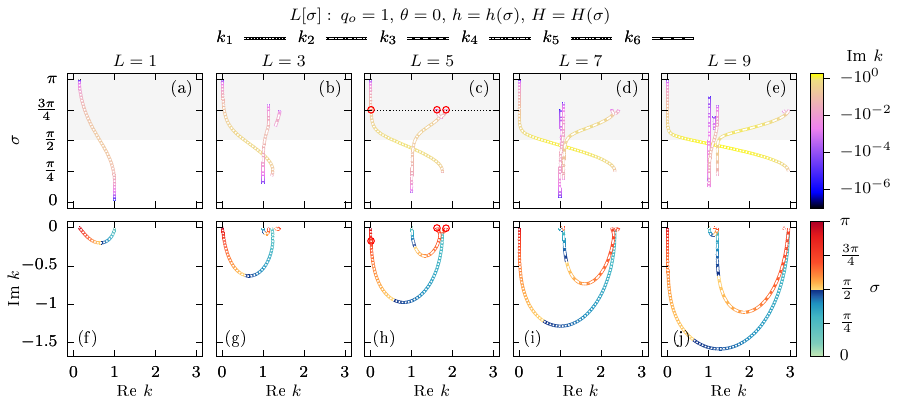}
    \caption{Zeros of $s_{11}(k)$ as a function of $\sigma$ for different values of $L$ in the full-box IP background configuration, with $q_o^2=h^2(\sigma)+H^2(\sigma)$ [see Eq.~\ref{eq:sigma}].
    The parameters $q_o=1$ and $\theta=0$ remain fixed.
    The upper row shows the location of the zeros in the $\Re k$-$\sigma$ plane whereas the bottom row shows the location of the zeros in the complex $k$ plane.
    The complex $k$-plane can be mapped onto Fig.~\ref{fig:DB_param_q} to retrieve the relevant physical information about the soliton solutions.
    The color coding shows the corresponding complementary quantity $\Im k$ (upper row) and $\sigma$ (bottom row).
    Only $\Re k_o>0$ are shown due to the symmetry of the zeros.
    The gray background in the top row panels corresponds to the equivalent case $h>0$ and $\alpha=\pi$.
    Red circles in (c) and (h) correspond to the zeros shown in Fig.~\ref{fig:GPE_IP_h}.
    Note that the quantities shown are measured in transverse oscillator units.}
    \label{fig:roots_LhH_th_0}
\end{figure*}
\begin{figure}[t]
    \centering
    \includegraphics[width=\columnwidth]{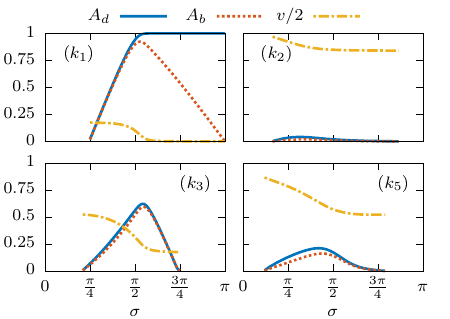}
    \caption{Amplitudes $A_d$, $A_b$, and velocity $v$ of $k_1$, $k_2$, $k_3$ and $k_5$ shown in Figs.~\ref{fig:roots_LhH_th_0}(e) and \ref{fig:roots_LhH_th_0}(j) (see legend) as a function of $\sigma$.
    Note that the quantities shown are measured in transverse oscillator units.
    }
    \label{fig:adabv2}
\end{figure}

We once more begin our investigation by exploring the spectra of an IP but full-box configuration.
Also here, $q_o=1$ and $\theta=0$ are held fixed, while $\sigma\in[0,\pi]$ and $L\in[1,9]$ are varied.
Recall that now $H(\sigma)\in[0,q_o]$ and $h(\sigma)\in[-q_o,q_o]$.
The corresponding spectrum of zeros is presented in Fig.~\ref{fig:roots_LhH_th_0}.
Since an IP configuration preserves symmetry, only the zeros of the pair with $\Re k_o>0$ are shown.

Let us first discuss the changes in the spectrum under an $L$ variation.
As in the zero-box configuration, increasing $L$ increases the number of zeros, and thus the soliton solutions.
This is readily seen in Figs.~\ref{fig:roots_LhH_th_0}(a)--(e) (top row).
In Fig.~\ref{fig:roots_LhH_th_0}(a), having $L=1$, $k_1$ is the only
pair of zeros, while for $L=5$ [Fig.~\ref{fig:roots_LhH_th_0}(c)]
already three different pairs of zeros, $k_1$, $k_2$ and $k_3$, are
potentially
present.
Finally, for $L=9$ [Fig.~\ref{fig:roots_LhH_th_0}(e)], up to six different pairs of zeros, $k_1,\dots,k_6$, occur.
Notice though that, in the latter panel, there is not a single value of $\sigma$ where all six solutions coexist at the same time. 
Moreover, most of the zeros at large $L$ values, i.e., $L=7$ and 9 [see Figs.~\ref{fig:roots_LhH_th_0}(d) and \ref{fig:roots_LhH_th_0}(e), respectively], remain around $\Re k_o = 1$, some of which having $|\Im k_o|<0.01$.
However, and in contrast to the zero-box configuration, the drops that characterized LIC solutions are absent in this setting.

Now, let us monitor the changes in the spectra as $\sigma$ increases.
Focusing initially on Figs.~\ref{fig:roots_LhH_th_0}(a)--\ref{fig:roots_LhH_th_0}(e) (top row) it is observed that, in contrast to the zero-box configuration, $\Re k_o$ does not always increase with $\sigma$.
This becomes apparent upon inspecting $k_1$, whose $\Re k_1$ always decreases as $\sigma$ increases.
Other examples are $k_2$ in Fig.~\ref{fig:roots_LhH_th_0}(d), or $k_4$ in Fig.~\ref{fig:roots_LhH_th_0}(e), as well as the bifurcation close to $\sigma=3\pi/4$, seen in Figs.~\ref{fig:roots_LhH_th_0}(c)--(e).
We also need to distinguish between the regimes $\sigma\in[0,\pi/2)$ (white background) and $\sigma\in(\pi/2,\pi]$ (gray background).
The former corresponds to $h>0$ (and $\alpha=0$), while the latter corresponds to $h<0$ or, equivalently, $h>0$ and $\alpha=\pi$ [see Eq.~\eqref{eq:initial_conditions}].

In the first regime ($\sigma<\pi/2$) $h$ decreases from $q_o$ to $0$, while $H$ increases from $0$ to $q_o$.
This implies that the system starts as a homogeneous condensate ($\sigma=0$) and, as $\sigma$ increases, the presence of the first component in the box decreases while the presence of the second component increases (see Fig.~\ref{fig:IC}).
Therefore, it is expected that no soliton solution emerges until the FBTC reaches certain conditions.
For example, in Fig.~\ref{fig:roots_LhH_th_0}(a) ($L=1$), $k_1$ is already present at very small values of $\sigma$.
This means that a small box is already enough to produce a soliton solution.
However, this soliton has a really low imaginary contribution, which means that the presence of the bright component is negligible.
Moreover, it is created at $\Re k_1 \approx 1$, which translates into a shallow ($A_d\approx 0$) fast moving ($v\approx c$) soliton (see Fig.~\ref{fig:DB_param_q}).
Then, as $\sigma$ increases further the box gets more and more filled by the second component, and thus the bright component of the ensuing DB soliton becomes dominant.
Note also that as $L$ increases $k_1$ emerges with larger $\Re k_1$, reaching almost $\Re k_1=3$ at $L=9$ [see Fig.~\ref{fig:roots_LhH_th_0}(e)]. 
It is also worth noticing that most of the zeros emerge around $\sigma=\pi/4$.
This is an important point, since $h(\sigma=\pi/4)=H(\sigma=\pi/4)=1/\sqrt{2}$.
Basically, it shows that the presence of the second component inside the box hinders the formation of soliton structures.
It is not until $h<H$ that the depth of the box is big enough to enhance the formation of DB solitons.
Additionally there exist also cases where zeros occur before $\sigma=\pi/4$.
However, these zeros have a low imaginary contribution and appear around $\Re k_o=1$ which, as stated above, corresponds predominantly to small disturbances moving with velocities proximal to the speed of sound.
Nevertheless, at $\sigma=\pi/2$ we recover the zeros from the zero-box configuration with $h=0$ and $H=q_o$.

On the contrary, in the second regime ($\sigma>\pi/2$) $|h|$ increases from $0$ to $q_o$ and $H$ decreases from $q_o$ to $0$.
Importantly here, $\alpha=\pi$, represents a situation where the first component presents a phase difference between the walls and the inside of the box [see Eq.~\eqref{eq:initial_conditions}].
Although this phase difference does not break the symmetry of the system, it introduces a constant perturbation in the system that needs to be taken into account, as we explain later on.
Similarly to the first regime, most of the zeros are present also here while $|h|<H$ ($\sigma<3\pi/4$).
Interestingly enough, in this regime $k_1$ exists for all $\sigma$, having $\Re k_1 \approx 0$ for a large range of $\sigma$, already from small $L$.
This is a direct cause of the phase difference $\alpha=\pi$, which forces the existence of at least one pair of solutions (see Ref.~\cite{Romero-Ros2021} and references therein).

To better understand the existence of all aforementioned zeros, we inspect Figs.~\ref{fig:roots_LhH_th_0}(f)--(j) (bottom row), which can be directly connected to the soliton characteristics shown in Fig.~\ref{fig:DB_param_q}.
In the complex $k$ plane, only the zeros with a high imaginary contribution are easily visible.
In this case, the most important difference with respect to the zero-box configuration is the parabolically-shaped trajectory of the zeros.

First, we focus on describing the zeros in Fig.~\ref{fig:roots_LhH_th_0}(j) with the aid of Fig.~\ref{fig:DB_param_q}.
Most of the zeros with a low imaginary contribution are merely dots around $\Re k_o = 1$ and $\Im k_o=0$, i.e., small-amplitude nearly sonic DB solitons with a negligible bright contribution.
On the other hand, $k_1$, $k_3$ and $k_5$ possess a higher imaginary contribution.
For instance, $k_3$ emerges at $\Re k_3 = 1.2$ and $\Im k_3 \approx 0$ already at $\sigma \approx \pi/4$.
DB solitons with $\Re k_o>1$ and $\Im k_o \approx 0$ are states that have extremely small amplitudes but large widths.
As an example, here the DB soliton corresponding to $k_3$ has
$A_d=0.006$ and $A_b=0.005$ at $\sigma=0.21\pi$ (when $k_3$ initially emerges) and a full width at half minimum for the dark
($\text{FWHM}_d$) and maximum for the bright component
($\text{FWHM}_b$) that read as $\text{FWHM}_d=560$ and
$\text{FWHM}_b=837$ (in H.O. units presented in Sec.~\ref{sec:model}).
Of course, such structures are practically impossible to be seen.
In addition, $k_3$ moves in this case with $v=c/2$.
Then, as $\sigma$ increases $|\Im k_3|$ rapidly increases and so do $A_d$ and $A_b$, which at the same time narrows the DB soliton.
Of course, bigger solitons move slower and $k_3$ is no exception.
Counterintuitively, the maximum bright contribution is found past $\sigma=\pi/2$, as indicated by the minimum of $k_3$ in Fig.~\ref{fig:roots_LhH_th_0}(j) at $\sigma=0.585\pi$.
Past this point, $\Im k_3$ starts to rapidly decrease, and so do $A_d$ and $A_b$, reaching $\Re k_3=3$ and $\Im k_3 \approx 0$ before ceasing to exist at $\sigma=3\pi/4$.
Recall that before disappearing, $k_3$ ends up again being a small wide DB soliton.
Also, note that $\Re k_3$ is always an increasing function of $\sigma$.
The trajectory of $k_3$ as $\sigma$ is varied can be better appreciated by inspecting Fig.~\ref{fig:adabv2}, where the mapping of $k_3$ onto Fig.~\ref{fig:DB_param_q} is shown along with further examples, i.e., $k_1$, $k_2$ and $k_5$.
Here, $A_d$, $A_b$ and $v$ are plotted against $\sigma$.
Interestingly, $v$ remains almost constant for most of the values of $\sigma$ in the second regime ($\sigma>\pi/2$), an outcome that is in turn related to the fact that $A_d\gtrapprox A_b$ in this regime.

A similar behavior to the $k_3$ one occurs also for $k_1$ within the first regime ($\sigma<\pi/2$).
Obviously, since $k_1$ and $k_3$ are on top of each other around $\Re k_o=3$ [see Fig.~\ref{fig:roots_LhH_th_0}(j)], when $k_1$ emerges for the first time it does so as a small and wide DB soliton.
Recall that different zeros never intersect, i.e., $k_i\neq k_j$.
Also note that $\Re k_1$ is a decreasing function with respect to a $\sigma$ variation. 
Then, as $\sigma$ increases, $\Re k_1$ decreases and $|\Im k_1|$ increases, which translates into larger $A_d$ and $A_b$, with $A_d\gtrapprox A_b$, and $v$ remaining almost constant.
Interestingly here, slightly before $\sigma=\pi/2$, $|\Im k_1|$ starts to decrease and both $\Re k_1$ and $\Im k_1$ rapidly approach $0$.
However, in this case only $A_b$ decreases as $\sigma$ keeps increasing.
On the contrary, $A_d \approx q_o$ and $v\approx 0$ independently of $\sigma$.
In Fig.~\ref{fig:adabv2} one can clearly discern the plateau of almost constant $v$ within the first regime and the constant values of $A_d$ and $v$ within the second regime.
The latter is a direct consequence of $\alpha=\pi$.
As discussed for the OP zero-box configuration, a phase difference of $\Delta\theta=\pi$ between two regions of a condensate will always lead to the formation of a static soliton whose dark component is a black soliton  with $A_d=q_o$ and $v=0$~\cite{Romero-Ros2021}.
Hence, once $\sigma>\pi/2$, then $h>0$ and $\alpha=\pi$ [see Eqs.~\eqref{eq:initial_conditions} and \eqref{eq:sigma}], which creates a phase jump at the edges that separate the inside of the box from its walls.
What is seen in Fig.~\ref{fig:adabv2} for $k_1$ at large values of $\sigma$ is a DB soliton formed by a black soliton and a bright counterpart that decreases as $\sigma$ increases ($H$ decreases).
Remarkably, it seems that in this case, when compared to the single-component scenario~\cite{Romero-Ros2021}, the presence of a second component does not affect the emergence of the black soliton but only that of the bright counterpart and the remaining solitons solutions.

\paragraph{\textbf{Out-of-phase background.}}

\begin{figure*}[t]
    \centering
    \includegraphics[width=\linewidth]{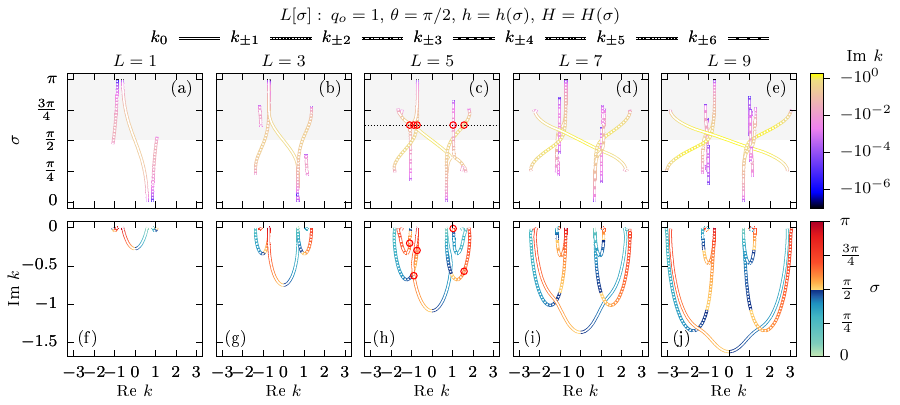}
    \caption{Zeros of $s_{11}(k)$ as a function of $\sigma$ for different values of $L$ in the full-box OP background configuration, with $q_o^2=h^2(\sigma)+H^2(\sigma)$ [see Eq.~\ref{eq:sigma}].
    The parameters $q_o=1$ and $\theta=\pi/2$ remain fixed.
    The upper row shows the location of the zeros in the $\Re k$-$\sigma$ plane whereas the bottom row shows the location of the zeros in the complex $k$-plane.
    The complex $k$-plane can be mapped onto Fig.~\ref{fig:DB_param_q} to retrieve the relevant physical information about the soliton solutions.
    The color coding shows the corresponding complementary quantity $\Im k$ (upper row) and $\sigma$ (bottom row).
    $k_+$ ($k_-$) corresponds to zeros with $\Re k_o>0$ ($\Re k_o<0$).
    $k_0$ is an unpaired solution.
    The gray background in the top row panels corresponds to the equivalent case $h>0$ and $\alpha=\pi$.
    Red circles in (c) and (h) correspond to the zeros shown in Fig.~\ref{fig:GPE_OP_h}.
    Note that the quantities shown are measured in transverse oscillator units.}
    \label{fig:roots_LhH_th_pi2}
\end{figure*}

Our last exploration of the spectra of zeros of $s_{11}(k)$ is performed for an OP full-box configuration.
Here, $q_o=1$ and $\theta=\pi/2$ are held fixed, with the latter
setting the two walls of the box out of phase.
Additionally, $L\in[1,9]$ and $\sigma\in[0,\pi]$ are varied.
Recall that $H(\sigma)\in[0,q_o]$ and $h(\sigma)\in[-q_o,q_o]$.
In this case the phase symmetry of the system is broken since $h\neq 0$ and $\theta\neq0$, and the system yields asymmetric solutions, with none of the zeros being paired for any value of $\sigma$.
The only exception here occurs for $\sigma=\pi/2$ (see below).
Therefore, in Fig.~\ref{fig:roots_LhH_th_pi2} we show the entire spectrum of zeros, i.e., $\Re k_o \in \mathbb{R}$.

It is worth noticing that given our particular choice of $\theta=\pi/2$ the zeros present an antisymmetry, evident in Figs.~\ref{fig:roots_LhH_th_pi2}(a)--(e), where the zeros are shown in the $\Re k$-$\sigma$ plane.
On the other hand, Figs.~\ref{fig:roots_LhH_th_pi2}(f)--(j) illustrate the zeros shown in the complex $k$ plane.
Here, the zeros are symmetric around $\Re k = 0$ and the antisymmetry is encoded in the color code introduced for the $\sigma$ variation.
The symmetry in the complex $k$-plane can be easily understood when looking back to Eq.~\eqref{eq:phase_jump}.
In the first regime $\sigma<\pi/2$, $\alpha=0$ and $\Delta\theta_\pm=\pm\theta$, while in the second regime, $\sigma>\pi/2$, $\alpha=\pi$, and $\Delta\theta_\pm=\mp\theta$.
This change of sign in $\Delta\theta$ implies a spatial reflection around $x=0$ [see Fig.~\ref{fig:IC}(a)] and gives rise to the (anti)symmetry of the spectra in a system with broken phase symmetry.
Therefore, we use the same line style to identify antisymmetric zeros, i.e., $\Re k_+>0$ and $\Re k_-<0$ (see legend in Fig.~\ref{fig:roots_LhH_th_pi2}), and in what follows we will comment only the zeros with $\Re k_o>0$.

As in the previous cases, increasing $L$ increases the number of zeros.
In Figs.~\ref{fig:roots_LhH_th_pi2}(a)--(e) (top row) the number of zeros increases from two, in Fig.~\ref{fig:roots_LhH_th_pi2}(a), to seven, in Fig.~\ref{fig:roots_LhH_th_pi2}(e).
Of course, in this case the number of zeros also depends on $\sigma$.
In general, increasing $\sigma$ while $\sigma<\pi/2$ (increasing $H$ and decreasing $h$) also increases the number of zeros.
On the other hand, increasing $\sigma$ while $\sigma>\pi/2$ (decreasing $H$ and increasing $h$) decreases the number of zeros.
Additionally, one needs to keep in mind that $\sigma>\pi/2$ also implies that $\alpha=\pi$ (with $h>0$).
It is also worth noticing that in some cases some of the zeros are present only in the first ($\sigma<\pi/2$) or in the second ($\sigma>\pi/2$) regime.
For instance, in Fig.~\ref{fig:roots_LhH_th_pi2}(e) $k_{+5}$ is found only for $\pi/4<\sigma<\pi/2$ (first regime).
Similarly, $k_{+2}$ and $k_{+6}$ are found only in the second regime.
The former appears for $\pi/2<\sigma<7\pi/8$ and the latter right before $\sigma=3\pi/4$.
Recall that this feature, i.e., all zeros do not coexist at the same time, was also found in the full-box IP case.
Yet another similarity with the IP case is that as $L$ increases, most of the zeros appear only between $\pi/4<\sigma<3\pi/4$ and are found mostly around $\Re k_o \approx \pm1$.

There are two peculiarities of the OP case also worth discussing.
The first one is the emergence of a DB soliton with a black soliton contribution, corresponding to $k_0$ at $\sigma=\pi/2$.
Notice that $k_0$ is the only unpaired zero and also the only zero bearing both positive and negative $\Re k_o$ values.
The change of sign, which is directly related to the velocity of the soliton [see Eq.~\eqref{eq:DB_parameters_gamma}], happens at $\sigma = \pi/2$ ($H=q_o$ and $h=0$), which coincides with the OP zero-box case discussed above (see Fig.~\ref{fig:roots_LH_q1_thpi2_h0_a0}).
In particular, at $\sigma=\pi/2$ we recover the solutions of the zero-box OP configuration.
The labeling of all $k_o$ is also kept accordingly.
The other peculiarity is found by $k_0$ and $k_{+1}$ in Figs.~\ref{fig:roots_LhH_th_pi2}(b) and \ref{fig:roots_LhH_th_pi2}(c), $k_{+1}$ and $k_{+3}$ in Fig.~\ref{fig:roots_LhH_th_pi2}(d), and $k_{+1}$ and $k_{+4}$ in Fig.~\ref{fig:roots_LhH_th_pi2}(e).
At low values of $H$, i.e., $\sigma\approx 0$ (or $\sigma \approx \pi$ for $k_{-}$), both zeros are almost on top of each other, implying that both solutions are almost identical, i.e., similar shape and velocity.
Additionally, locally both edges of the box ($x=\pm L$) are equivalent, a situation  more pronounced as $H\rightarrow 0$ and $h\rightarrow \pm q_0$, which reduces to the single-component case.
Basically, the formation of such similar solutions is a direct consequence of our choice of parameters which define an equivalent phase-jump at both edges of the box, $\Delta\theta_-=\Delta\theta_+$ [see Eq.~\eqref{eq:phase_jump}].

Figures \ref{fig:roots_LhH_th_pi2}(f)--\ref{fig:roots_LhH_th_pi2}(j) show the solutions in the complex $k$ plane.
Most of the properties for this representation are already mentioned in the IP case, whose zeros look alike.
Yet, in this case, we were able to identify the only case where the maximum bright soliton contribution of a particular soliton solution coincides with the maximum presence of the second component in the box.
Of course, this zero is $k_0$ and the maximum contribution of its bright component occurs at $\sigma=\pi/2$, precisely when the solution is the static DB soliton.

\subsection{Nucleation of DB soliton trains: Without confinement} \label{sec:homogeneous}

In this section we intend to verify the analytical results captured by the discrete eigenvalues identified in Section~\ref{sec:analytic_solutions}.
Initially, we numerically solve the CGPE [Eq.~\eqref{eq:CGPE_adim}] in the absence of a trapping potential, i.e., $\Omega=0$,  by employing a fourth-order Runge-Kutta integrator accompanied by a second order finite-differences method accounting for the spatial derivatives.
The spatial and temporal discretizations are $dx = 0.1$ and $dt = 0.001$, respectively, while the domain of integration used is located at $|x|=2500$ so as to avoid finite size effects, for the times of interest herein.
In the following, we fix $L=5$ and $q_o=1$, while $\theta=\{0,\pi/2\}$ for both the zero- and the full-box configurations.

Below we present our findings regarding the dynamical nucleation of DB solitons via the matter-wave interference method of two condensates in the presence of a second species in-between, also featuring the counterflow (see Fig.~\ref{fig:IC}).
It is important to note that the various DB solitons nucleated when utilizing the initial condition ansatz of Eq.~\eqref{eq:initial_conditions} have finite velocities and, in general, interact with each other.
Therefore, the analytical findings can be compared to the numerical
ones only in the asymptotic limit $t\to\infty$.
In this limit, each DB soliton can be considered well separated and independent from the rest of the solitary waves.
In this sense, discrepancies between the analytical DB soliton solutions of Eq.~\eqref{eq:DB_solution} and the numerically formed ones are expected to decrease as $t\to\infty$, as it is found and discussed later on.
Finally, in the results to be presented below, the analytical DB soliton solution is centered at $x_0=0$, unless stated otherwise.

\subsubsection{\textbf{Zero-box configuration}} \label{sec:zero-box_homo}

\begin{figure}[t]
    \centering
    \includegraphics[width=\columnwidth]{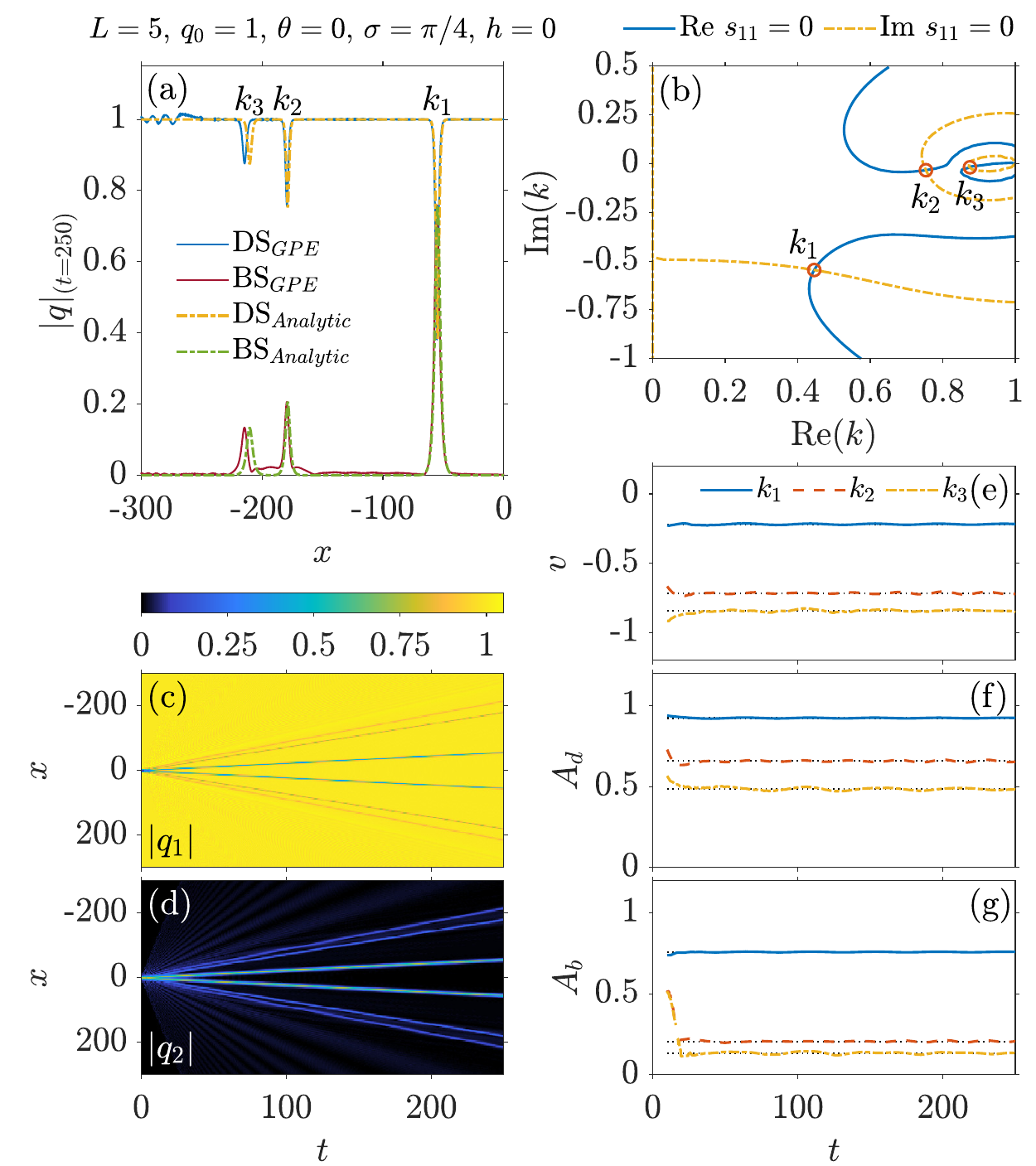}
    \caption{Dark-bright soliton solutions stemming from a zero-box configuration with an in-phase background having $L=5$, $q_o=1$, $\theta=0$, $\sigma=\pi/4$, and $h=0$ [cf. Figs.~\ref{fig:roots_LH_q1_th0_h0_a0}(c) and \ref{fig:roots_LH_q1_th0_h0_a0}(h)].
    (a) Snapshot of $|q|$ at $t=250$ given by the CGPE \eqref{eq:CGPE_adim} (solid lines) and the analytical solutions \eqref{eq:DB_solution} (dotted-dashed lines), for both dark (DS) and bright (BS) soliton counterparts.
    (b) Contour plot of $\Re s_{11}=0$ (solid blue line) and $\Im s_{11}=0$ (dashed yellow line) on the complex $k$-plane.
    The zeros, $k_o$, are depicted with red circles and only the zeros of each pair with $\Re k_o >0$ are shown.
    The labeling of zeros is that of Fig.~\ref{fig:roots_LH_q1_th0_h0_a0}, with
    $k_1=0.4456-i0.5455$, $k_2=0.7535-i0.0339$, and $k_3=0.8751-i0.0189$.
    (c), (d) Spatiotemporal evolution of the dark, $|q_1|$, and bright, $|q_2|$, soliton components.
   	Temporal evolution of (e) the instantaneous velocity, $v$, and (f) the dark, $A_d$, and (g) bright, $A_b$, soliton amplitudes.
    The corresponding asymptotic values are depicted with dotted black lines.
    Note that the quantities shown are measured in transverse oscillator units.
    }
    \label{fig:GPE_IP}
\end{figure}

\paragraph{\textbf{In-phase background.}}
Our first result is presented in Fig.~\ref{fig:GPE_IP}.
It corresponds to the zero-box configuration ($h=0$) with an IP background ($\theta=0$).
Here, we have chosen $\sigma=\pi/4$ as a representative example.
The zeros of this particular initial configuration are shown in Figs.~\ref{fig:roots_LH_q1_th0_h0_a0}(c) and \ref{fig:roots_LH_q1_th0_h0_a0}(h), pinpointed with red circles.
In particular, three pairs of DB solitons are predicted by our analytical method and indeed found in the dynamical process.
For instance, in Fig.~\ref{fig:GPE_IP}(a) the norm of the wave function, $|q|$, of each component at $t=250$ is shown, and all three pairs of DB solitons are clearly formed.
Note that due to the symmetry of the solutions, only the left moving solitons $v<0$ are illustrated.
The same holds for their corresponding zeros shown in Fig.~\ref{fig:GPE_IP}(b), where only the pair with $\Re k_o>0$ is depicted.
In particular, Fig.~\ref{fig:GPE_IP}(b) is equivalent to Fig.~\ref{fig:roots_LH_q1_th0_h0_a0}(h), as can be inferred from the location of the zeros in the complex $k$ plane. 
Notice that for consistency the notation introduced here follows that of Fig.~\ref{fig:roots_LH_q1_th0_h0_a0}.
 
A remarkably good agreement between the analytical estimates and the numerically formed DB solitons occurs already at $t=250$ (see Fig.~\ref{fig:GPE_IP}).
Particularly, both the numerically found solutions (solid lines) and
the analytically obtained ones (dotted-dashed lines) fall almost
perfectly on top of each other.
This also confirms the validity of the numerical scheme, given the
exact
nature of the IST analysis at the level of the integrable Manakov model.
The major discrepancy observed in this case corresponds to the shallower and faster DB soliton solution $k_3$.
There exist mainly three different sources that can give rise to such a discrepancy:
(i) as previously discussed, one should only expect both solutions to
exactly coincide at $t\to\infty$ or, equivalently, for such traveling
solutions to $x\to\pm\infty$.
Yet, the bright solitons of the $k_2$ and $k_3$ solutions still bear a finite background reminiscent of the filling of the box in the initial configuration.
We attribute the presence of this background to the intercomponent interaction, an effect which is enhanced for initially overlapping components, as will be shown in the full-box configuration results;
(ii) $k_3$ is the fastest DB soliton, which implies that $k_3$ is the wave that remains for longer times coupled to the emitted radiation, some of which is still visible around $x\approx 300$.
This effect is enhanced the faster the soliton is;
(iii) the interaction between the $k_2$ and $k_3$ DB solitons may play
a role, since both waves travel close to each other for a reasonable
long amount of time.
Indeed, Fig.~\ref{fig:GPE_IP}(c) [\ref{fig:GPE_IP}(d)] shows the spatiotemporal evolution of the wave function $|q_1|$ [$|q_2|$], which hosts the dark [bright] counterpart of the DB solitons in question.
Here, it is clear that $k_2$ and $k_3$, namely the outermost traveling DB solitons, remain close to each other during evolution.

Next, in order to extract the DB soliton characteristics, we
numerically follow the center of mass (c.m.) of each DB soliton, i.e.,
$x_{CM}=\qty(\int_{x_l}^{x_r}x|q|^2\dd
x)/\qty(\int_{x_l}^{x_r}|q|^2\dd x)$ with $x_{l,r}$ defining the
integration limits around each dark soliton core.
This also provides access to their instantaneous velocity, $v=\dd x_{CM}/\dd t$.
To obtain the c.m., we trace the dark soliton minima. 
From the position of the latter, we consecutively extract the dark, $A_d$, and bright, $A_b$, soliton amplitudes, and compare them with their corresponding asymptotic analytical values [Eq.~\eqref{eq:DB_parameters}].
$v$, $A_d$, and $A_b$ are depicted in Figs.~\ref{fig:GPE_IP}(e)--\ref{fig:GPE_IP}(g), respectively, for $t>10$ since at the very beginning of the dynamics it is not possible to identify any individual solitonic structure.
In all cases, it becomes apparent that the numerical predictions approach the analytical estimates (dotted black lines) as $t\to\infty$.
Notice also the small-amplitude oscillations performed by $v$, $A_d$, and $A_b$ around their asymptotic value, attributed to the counterflow process that leads to the soliton formation.
\begin{figure}[t]
    \centering
    \includegraphics[width=\columnwidth]{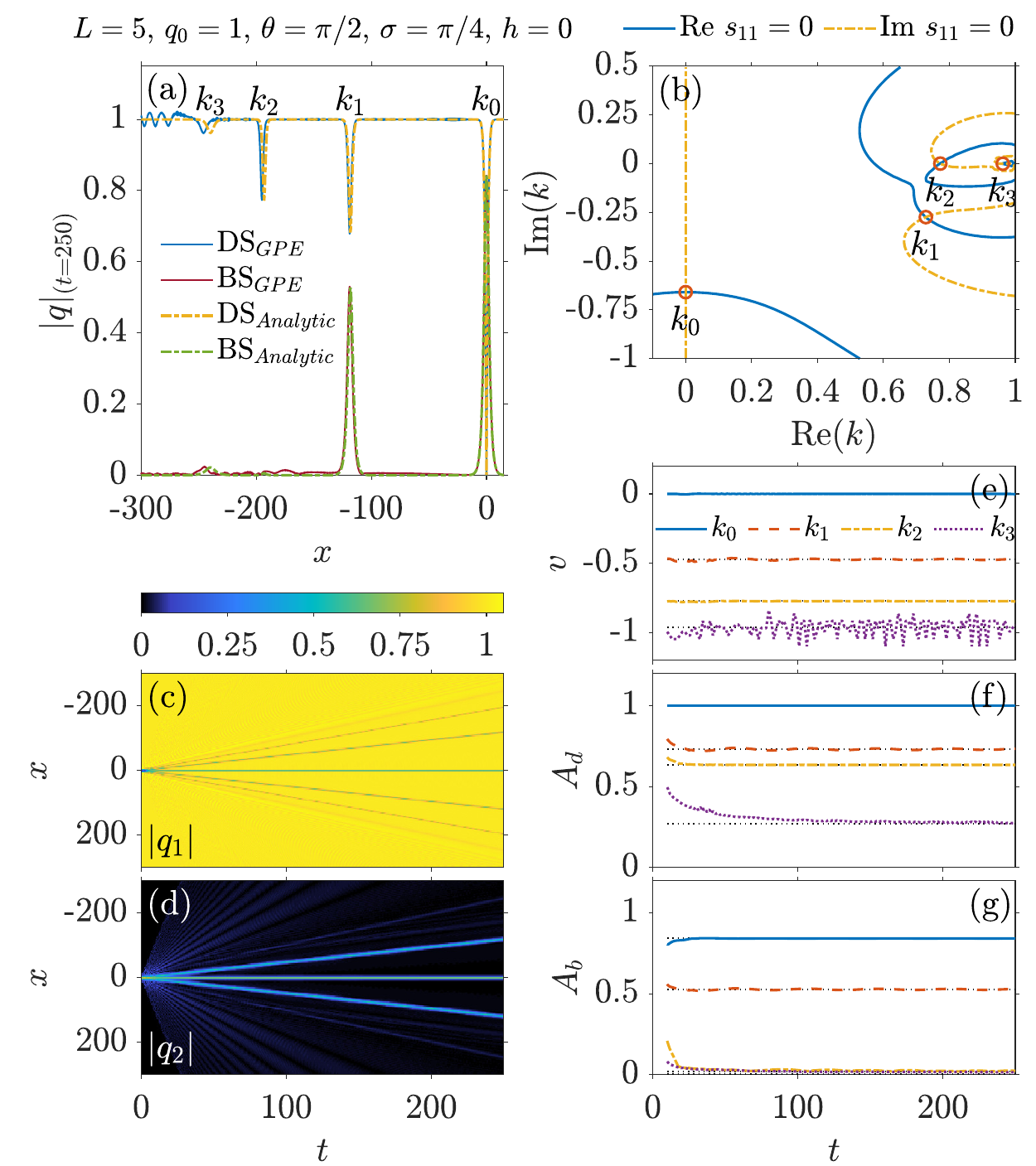}
    \caption{Same as Fig.~\ref{fig:GPE_IP} but for $L=5$, $q_o=1$, $\theta=\pi/2$, $\sigma=\pi/4$, and $h=0$.
    This initial configuration corresponds to a zero-box configuration with an out-of-phase background [cf. Figs.~\ref{fig:roots_LH_q1_thpi2_h0_a0}(c) and \ref{fig:roots_LH_q1_thpi2_h0_a0}(h)].
    The labeling of zeros is that of Fig.~\ref{fig:roots_LH_q1_thpi2_h0_a0} with $k_0=-i0.6590$, $k_1=0.7287-i0.2744$, $k_2=0.7726-i2.5\times10^{-5}$, and $k_3=0.9622-i8.8\times10^{-4}$.
    Note that the quantities shown are measured in transverse oscillator units.
    }
    \label{fig:GPE_OP}
\end{figure}

\paragraph{\textbf{Out-of-phase background.}}
Now, we present the results for a zero-box configuration ($h=0$) but with an OP background ($\theta=\pi/2$).
The zeros of this initial configuration were presented in Figs.~\ref{fig:roots_LH_q1_thpi2_h0_a0}(c) and \ref{fig:roots_LH_q1_thpi2_h0_a0}(h), and we have chosen $\sigma=\pi/4$ as the most relevant case for this particular set of parameters.
Our analytics predict, in this case, four zeros: a static unpaired DB soliton and three pairs of DB solitons.
Such solutions are marked with red dots in Figs.~\ref{fig:roots_LH_q1_thpi2_h0_a0}(c) and \ref{fig:roots_LH_q1_thpi2_h0_a0}(h), and are also shown in Fig.~\ref{fig:GPE_OP}(b).
Note that once more, the solutions are symmetric with respect to the origin ($x=0$) and for clarity we only show those with $\Re k_o>0$.
Each of the zeros illustrated in Fig.~\ref{fig:GPE_OP}(b) corresponds to a particular DB soliton solution, shown in Fig.~\ref{fig:GPE_OP}(a).
Again, the numerically observed waveforms (solid lines), obtained upon solving the CGPE with this particular OP zero-box configuration, fall on top of the analytical solutions (dotted-dashed lines) given by the zeros shown in Fig.~\ref{fig:GPE_OP}(b).
As in the IP zero-box configuration, we find also here that $k_3$ is again the DB soliton that presents the larger deviation from its analytical state.
Nevertheless, this OP case features two interesting structures not seen in the IP case.
The first one is the occurence of a static DB soliton, $k_0$, located at $x=0$.
As we discussed in Section~\ref{sec:analytic_solutions}, an OP configuration allows the formation of static DB solitons consisting of a black soliton ($v=0$) and its symbiotic bright counterpart.
The second one is related to the soliton $k_2=0.7726-i2.5\times10^{-5}$, which possesses an almost negligible imaginary contribution.
Recalling our discussion of Sec.~\ref{sec:analytic_solutions}, the bright counterpart of a DB soliton solution is mostly defined by the imaginary contribution of its corresponding zero.
Therefore, since in this case $\Im k_2\sim 10^{-5}$ we expect and indeed confirm the formation solely of a dark soliton.
Notice however the minuscule second component contribution that is in turn related, as in the IP case, to a small background reminiscent of the interaction between the two components during the dynamics.
Similarly, $k_3$ with $\Im k_3\sim 10^{-4}$ can also be
practically treated as a dark soliton.

In Figs.~\ref{fig:GPE_OP}(c) and \ref{fig:GPE_OP}(d), the spatiotemporal evolution of $|q_1|$ and $|q_2|$, respectively, clearly shows a static DB soliton at $x=0$ and the three pairs of DB solitons moving outwards.
Note that, in Fig.~\ref{fig:GPE_OP}(d), the bright component of $k_2$ is not seen and the bright component of $k_3$ is barely visible.
Figures~\ref{fig:GPE_OP}(e)--\ref{fig:GPE_OP}(g) demonstrate the evolution of the numerically obtained $v$, $A_d$, and $A_b$, respectively, together with their asymptotic analytical values (dotted black lines).
Yet, again, the numerical quantities asymptotically approach their corresponding analytical values.
In this case, small-amplitude oscillations in $v$, $A_d$, and $A_b$, caused by the dynamical formation of the solitonic entities are also found, [cf. $k_1$ in Figs.~\ref{fig:GPE_OP}(e)--\ref{fig:GPE_OP}(g)].
In contrast, the velocity of $k_3$, the fastest DB soliton, features abrupt and irregular oscillations.
This is due to the fact that we are computing the instantaneous velocity, $v=\dd x_{c.m.}/\dd t$, by integrating around each dark soliton core.
A closer inspection of Fig.~\ref{fig:GPE_OP}(a) reveals that some noise is still present around the DB structure at $t=250$.
Since this noise is not constant, when calculating $x_{c.m.}$ small irregular changes lead to the irregular oscillations observed in $v$.

\subsubsection{\textbf{Full-box configuration}} \label{sec:full-box_homo}

\begin{figure}[t]
    \centering
    \includegraphics[width=\columnwidth]{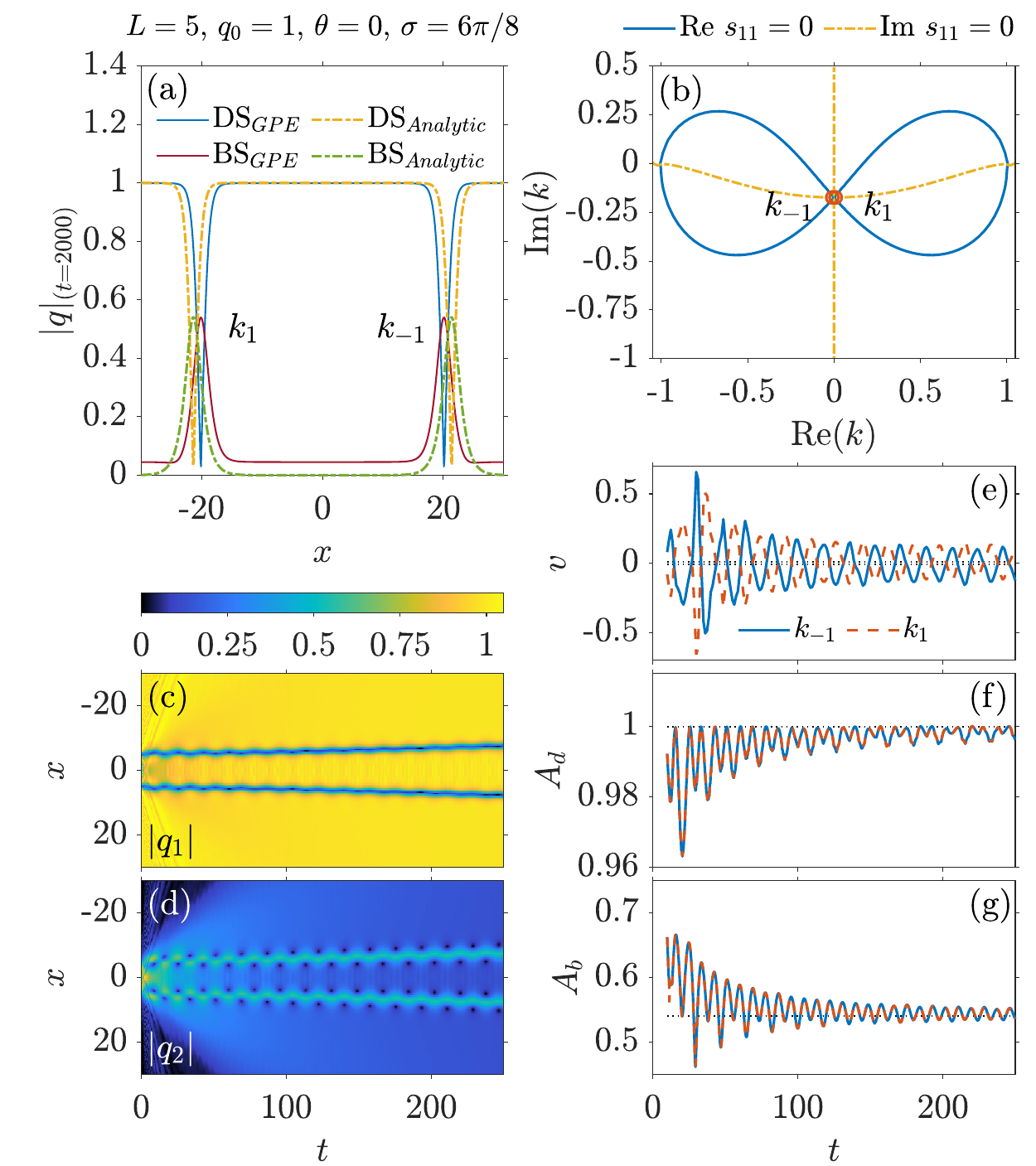}
    \caption{Same as Fig.~\ref{fig:GPE_IP} but for $L=5$, $q_o=1$, $\theta=0$, $\sigma=6\pi/8$.
    This initial configuration corresponds to a full-box configuration with an in-phase background [cf. Figs.~\ref{fig:roots_LhH_th_0}(c) and \ref{fig:roots_LhH_th_0}(h) ].
    In this case the only relevant solutions are $k_{\pm1}=\pm0.0098-i0.1737$.
    We omitted $k_{\pm2}=\pm1.6237-i0.0060$ and $k_{\pm3}=\pm1.8440-i0.0080$ (see text). 
    Note the long-time dynamics in (a).
    Note also that the quantities shown are measured in transverse oscillator units.
    }
    \label{fig:GPE_IP_h}
\end{figure}

\paragraph{\textbf{In-phase background.}}
In the full-box configuration, the center of the box is fully filled, i.e., $q_o^2=h^2(\sigma)+H^2(\sigma)$ for all values of $\sigma$ [see Eq.~\eqref{eq:sigma}].
Initially, we explore the IP background ($\theta=0$) upon choosing $\sigma=6\pi/8$.
This in turn implies that the first component inside the box is OP with respect to the two sides of the box [see Eq.~\eqref{eq:phase_jump}].
The analytical solutions for this particular choice of parameters were presented in Figs.~\ref{fig:roots_LhH_th_0}(c) and \ref{fig:roots_LhH_th_0}(h), with the relevant zeros being marked by red dots.
In total, three pairs of DB soliton solutions are found.
However, in what follows we only discuss the pair $k_{\pm1}$.
The other two pairs of solutions correspond to DB solitons with FWHM~$\gtrsim 10^2$ and amplitudes $A_{d,b}\lesssim 10^{-3}$ (see also the discussion in Sec.~\ref{sec:analytic_solutions}), and thus we omit them.

Illustrated in Fig.~\ref{fig:GPE_IP_h}(b) are the zeros $k_{\pm1}=\pm0.0098$~-~$i0.1737$, which lie almost on top of each other since $\Re k_{\pm1}\approx 0$.
In Fig.~\ref{fig:GPE_IP_h}(a) we compare the numerically found DB solitons (solid lines), stemming from the CGPE, with the analytical ones (dotted-dashed lines), obtained using our analytical tools presented in Sec.~\ref{sec:model}.
Although, in this case, we show the DB soliton profiles at later evolution times ($t=2000$), the numerical solutions do not completely coincide yet with the analytical ones.
The reason why this happens is not only that our analytical method provides solutions at $x\to\pm\infty$ or, equivalently, at $t\to\infty$, but also the interaction between the pair of DB solitons at early times.
Additionally, note that at these earlier times, shown in Figs.~\ref{fig:GPE_IP_h}(c) and \ref{fig:GPE_IP_h}(d), the pair of DB solitons does not emerge at $x_o=0$ but at $x_o=\pm5$ [see Eqs.~\eqref{eq:DB_solution}].
As discussed in Sec.~\ref{sec:full_box_analytic}, the phase-jump $\Delta\theta_\pm=\pi$ in the first component between the inner and the outer sides of the box leads to the formation of a pair of (almost) black-bright solitons where the phase-jump takes place.
Moreover, the latter implies $v\sim0$, which enhances the interaction between the pair of DB solitons for longer times than in the previously discussed cases, as mentioned before.

However, despite the fact that we cannot properly capture the early stages of the dynamics for these pairs of DB solitons, an interesting observation, absent in the previous explorations, can be made.
For instance, during the early dynamics, the presence of a non-negligible background in the minority species radically changes the behavior of a typical DB soliton, and our numerically identified waveforms morph into beating DB solitons~\cite{Hoefer2011,Yan2012}.
Indeed, the spatiotemporal evolution of both the dark and the bright soliton components [see Figs.~\ref{fig:GPE_IP_h}(c) and \ref{fig:GPE_IP_h}(d), respectively] reveal the characteristic beating of such solitonic entities.
Importantly, these beating solitons, however, are not ``discernible'' at the level of the eigenvalues of the IST analysis.
Here, we want to point out that the bright solitons of the DB entity $k_{\pm1}$ are in phase and therefore the DB solitons interaction is repulsive~\cite{Segev1998}, an effect that can be discerned by closely inspecting Figs.~\ref{fig:GPE_IP_h}(c) and \ref{fig:GPE_IP_h}(d) at later times.

Now, let us discuss Fig.~\ref{fig:GPE_IP_h}(e) showcasing $v$.
Since $\Delta\theta_\pm=\pi$, and thus $\Re k_{\pm1}\approx 0$ (see Sec.~\ref{sec:full_box_analytic}), the analytic velocities of such solitons are close to zero.
Also, since $k_{\pm1}$ are a pair, their velocities have opposite signs.
However, the interesting phenomenon found here is the beating performed by the DB soliton pair due to the presence of the finite background in the second component.
Indeed, here we can clearly see how $v$ oscillates while asymptotically approaching its analytical value, and that $v$ undergoes damped oscillations while approaching its asymptotic value.
The damping behavior is inherently related to a progressive decrease of the finite background over time.
In order to reach their asymptotic velocities, one should wait for the finite background of the second component to vanish and for the solitons to be well separated from each other to avoid interacting.
The same applies to the dark and bright amplitudes, shown in Figs.~\ref{fig:GPE_IP_h}(f) and \ref{fig:GPE_IP_h}(g), respectively.
Nonetheless, Figs.~\ref{fig:GPE_IP_h}(f) and \ref{fig:GPE_IP_h}(g)
provide a visual confirmation of the symmetry of the solutions, where
solitons undergo the same amplitude oscillations, the latter being
also a characteristic of beating DB solitons~\cite{Yan2012} [see the discussion around Eq.~\eqref{eq:beating_Yan}].
\begin{figure}[t]
    \centering
    \includegraphics[width=\columnwidth]{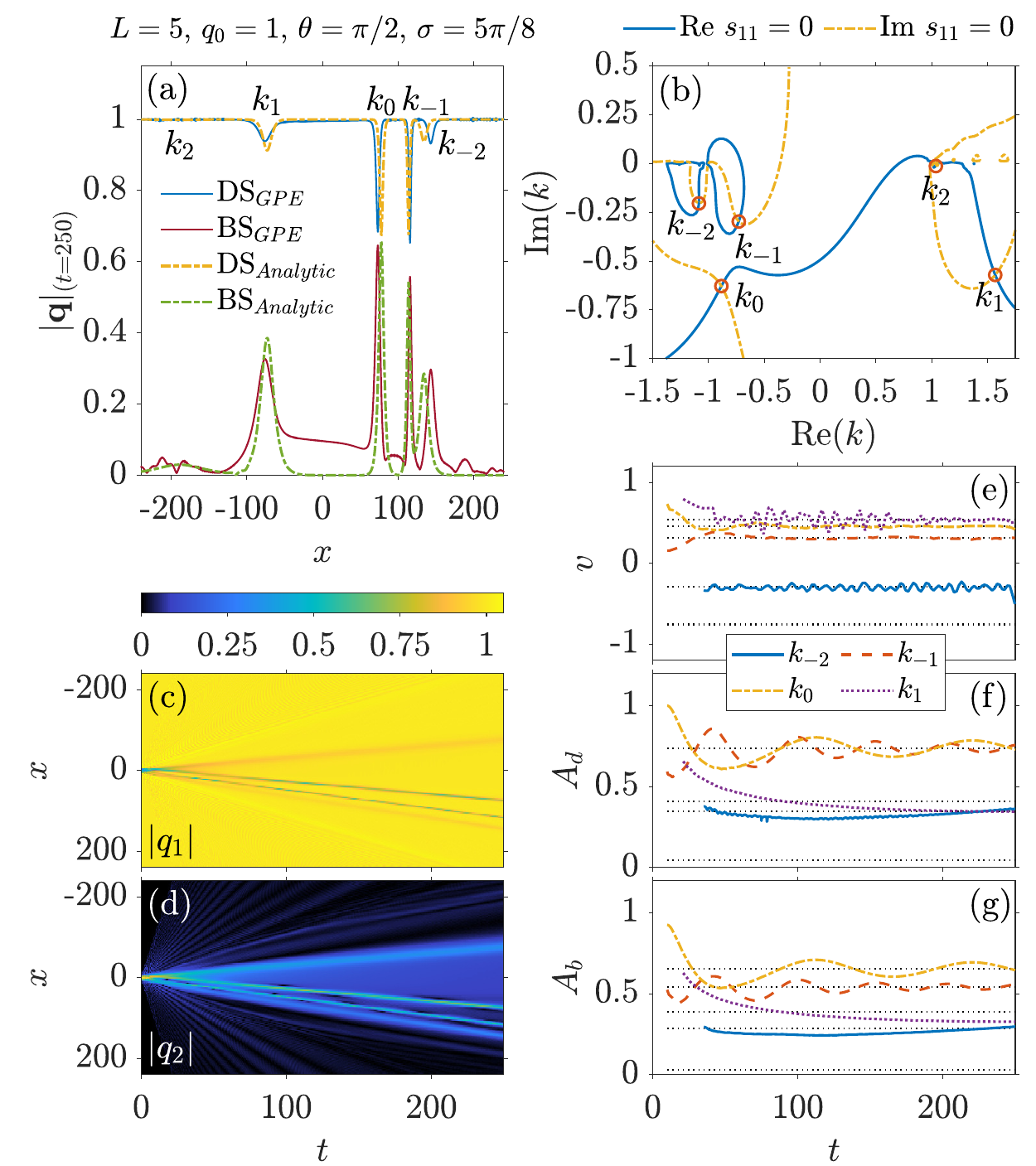}
    \caption{Same as Fig.~\ref{fig:GPE_IP} but for $L=5$, $q_o=1$, $\theta=\pi/2$, $\sigma=5/\pi8$.
    This initial configuration corresponds to a full-box configuration with an out-of-phase background [cf. Fig.~\ref{fig:roots_LhH_th_pi2}(c) and \ref{fig:roots_LhH_th_pi2}(h)].
    In this case the zeros are not symmetric.
    The labeling of zeros is that of Fig.~\ref{fig:roots_LhH_th_pi2} with
    $k_{-2}=-1.0858-i0.2038$, $k_{-1}=-0.7285-i0.29769$, $k_0=-0.8843-i0.6277$, $k_1=1.5701-i0.5708$ and $k_2=1.0381-i0.0127$.
    Note that the quantities shown are measured in transverse oscillator units.
    }
    \label{fig:GPE_OP_h}
\end{figure}

\paragraph{\textbf{Out-of-phase background.}}

The last parametric selection consists on a full-box configuration, i.e., $q_o^2=h^2(\sigma)+H^2(\sigma)$ [see eq.~\eqref{eq:sigma}], with an OP background ($\theta=\pi/2$).
The analytical solutions for such an initial configuration were shown in Figs.~\ref{fig:roots_LhH_th_pi2}(c) and \ref{fig:roots_LhH_th_pi2}(h).
Here, we choose as a case example $\sigma=5\pi/8$, with the relevant zeros pinpointed with red dots.

In Fig.~\ref{fig:GPE_OP_h}(b) the five zeros corresponding to this particular initial configuration are depicted with a red circle.
In Fig.~\ref{fig:GPE_OP_h}(a), the analytical solutions obtained using these zeros (dotted-dashed lines) are compared to the numerical solutions (solid lines), obtained by solving the CGPE.
Both solutions almost fall on top of each other.
Most of the discrepancies found here can be attributed as in the preceding sections to the presence of a finite background, as well as DB-DB soliton interactions.
In Fig.~\ref{fig:GPE_OP_h}(a), the most extreme case is that of $k_2=1.0381$~-~$i0.0127$, where the dark component of the DB soliton cannot be identified.
This is a direct consequence of the fact that $\Re k_2\approx1$, as discussed in Sec.~\ref{sec:full_box_analytic}.
Additionally, the corresponding bright part of $k_2$ is disturbed by the spreading of the finite background.

The spatiotemporal evolution of the dark  and bright soliton components [see Figs.~\ref{fig:GPE_OP_h}(c) and \ref{fig:GPE_OP_h}(d), respectively] demonstrates the asymmetric nature of the ensuing DB waves for this parametric selection.
Of course, $k_2$ is not discernible in Fig.~\ref{fig:GPE_OP_h}(c), while in Fig.~\ref{fig:GPE_OP_h}(d) the finite background on top of which the bright solitons are formed is clearly visible.
Among them, $k_0$ and $k_{-1}$ are seen to undergo small-amplitude oscillations, resembling beating DB solitons. 
Unfortunately, the oscillations around their c.m. are not pronounced enough so as to be captured by the temporal evolution of the instantaneous velocity in Fig.~\ref{fig:GPE_OP_h}(e).
Nevertheless, we are still able to follow the c.m. of most of the evolved solitonic entities, showcasing this way that they approach their asymptotic analytical values (dotted black lines) as $t\to\infty$.
The only exception here is the nearly sonic $k_2$ soliton, whose CM cannot be separated from the surrounding radiation.
Yet, we left its analytical value as a reference.
Figures~\ref{fig:GPE_OP_h}(f) and \ref{fig:GPE_OP_h}(g) illustrate the evolution of $A_d$ and $A_b$ for each DB soliton formed.
Noteworthy here is the damping behavior of $A_d$ and $A_b$ associated with the beating solitons $k_0$ and $k_{-1}$. 
Finally, it is worth commenting $k_{-2}$ is still far below its asymptotic value, while $k_1$ closely approaches its asymptotic value from above around $t=250$.

\subsection{Nucleation of DB soliton trains: With confinement} \label{sec:trap}

In BEC experiments, harmonic confinement is naturally introduced.
For this reason, in this section we aim to generalize our findings in the presence of a harmonic trapping potential and, for the numerical considerations to be presented below, we 
turn on the trapping potential in Eq.~\eqref{eq:CGPE_adim}.
Hereafter, we fix $\Omega=0.011$.
As in Sec.~\ref{sec:homogeneous}, we will first present the results for the zero-box configuration, and the results for the full-box configuration will follow.

Before proceeding to the results, first we want to remark that in the presence of a harmonic confinement our analytical estimates, obtained by solving the direct scattering problem (see Sec.~\ref{sec:model}), are not expected to provide valid solutions.
For example, we assumed NZBC which in turn define the asymptotic behavior of the solitons formed in terms of velocity and amplitude.
It is clear that in the presence of the harmonic potential such NZBC cannot be fulfilled.
However, with an appropriate choice of parameters, the analytical solutions of the untrapped scenario (see Sec.~\ref{sec:homogeneous}) can be used as approximate solutions for the trapped scenario as we shall later show.
For instance, our choice of a wide trapping potential ($\Omega=0.011$) provides a ground state of the first component flatter around the center of the trap, which can at least locally resemble a constant background like that of the homogeneous case.

To induce the dynamics in our system, we first find the ground state of a single-component BEC by means of imaginary-time propagation.
Then, we embed on top of the ground state our initial configuration [see Eq.~\eqref{eq:initial_conditions}].
A schematic illustration of the aforementioned initial state is provided in Fig.~\ref{fig:IC}(b).
Moreover, to offer a direct comparison between the untrapped and the trapped scenarios, our choice of parameters is the same as in Sec.~\ref{sec:homogeneous}, i.e., $L=5$, $q_o=1$, $\sigma={\pi/4}$ with $\theta=\{0,\pi/2\}$, and $\sigma=\{6\pi/8,5\pi/8\}$ with $\theta=\{0,\pi/2\}$, respectively (see also the relevant discussion around Figs.~\ref{fig:GPE_IP}--\ref{fig:GPE_OP_h}).

In order to characterize the solutions, we compute in each case the
oscillation frequency of the DB solitons using the following,
well-established
expressions \cite{Busch2001} (see also, e.g.,~\cite{Kevrekidis2016}):
\begin{subequations} \label{eq:osc_freq}
\begin{gather}
    \omega_o^2 = \Omega^2\qty(\frac{1}{2}-\frac{\chi}{\chi_o})\,,
    \label{eq:omega_o}
    \\
    \chi_o=8\sqrt{1+\qty(\frac{\chi}{4})^2}\,, \qquad 
    \chi\equiv\frac{N_b}{q_o}\,, 
    \\
    N_b \equiv \int_{-\infty}^\infty |q_b(x,t)|^2 \dd x = 2\qty(\frac{q_o^2}{|z_o|^2}-1)\Im z_o\,, 
    \label{eq:Nb}
\end{gather}
\end{subequations}
and describe the motion of the center of the DB solitons as
\begin{align}
	x_c(t) = \frac{v_o}{\omega_o}\sin(\omega_o t + \phi_o) + x_o \,,
	\label{eq:trap_trajectory}
\end{align}
Here, the amplitude of the oscillation is related to the velocity of the DB solitons [see Eq.~\eqref{eq:v}] and the frequency of the trap [see Eq.~\eqref{eq:omega_o}].
Additionally, $x_o$ is the equilibrium position, and $\phi_o$ is an additional phase factor.
Both ${x}_o$ and $\phi_o$ are fixed to zero unless stated otherwise.

It is important to remark here that, contrary to the single-component dynamics of dark and bright solitons in the presence of a harmonic potential, the amplitudes of each dark and bright counterpart of a DB soliton are not constant over time, but oscillate.
Hence, we propose the following DB soliton estimate accounting for the amplitudes' dynamics (see Appendix~\ref{app:amplitudes}):
\begin{subequations} \label{eq:DB_solution_trap}
\begin{gather}
	q_d^{(n)}(x,t)= q_o\cos{\beta_n(t)}-iq_o\sin{\beta_n(t)}\tanh{\big[\nu(t)(x-x_c(t))\big]} \,,
    \label{eq:dark_solution_trap}
    \\
    q_b^{(n)}(x,t)= -i\sin{\beta_n(t)} \sqrt{q_o^2-\abs{z_n}^2}\sech{\big[\nu(t)(x-x_c(t))\big]} \,, 
    \label{eq:bright_solution_trap}
\end{gather}
\end{subequations}
where we found that the angle parameter is now time dependent with the form
\begin{align}
\cos^2\beta_n(t) = \cos^2\beta_n\cos^2(\omega_o t)+\frac{1}{2q_o^2}\Omega^2\qty(\frac{v_n}{\omega_o})^2\sin^2(\omega_o t) \,.
\label{eq:beta_t}
\end{align}
From here, the uniformization variable can be expressed as $z(t) =|z_o|e^{i\beta(t)}$.
The other time dependent parameters can be obtained by substituting Eq.~\eqref{eq:beta_t} in Eq.~\eqref{eq:xi_nu}.
Of course, if we turn off the trap ($\Omega=0$ and $\omega_o=0$) we recover $\beta(t)=\beta_o$.

Last, we design in-trap analytical estimates of the dark  and bright soliton solutions as follows,
\begin{subequations} \label{eq:DB_solitons_trap}
	\begin{gather}
	|q_1(x,t)|^2 = \abs{ q_o^2\abs{\prod_n \frac{q_{d}^{(n)}(x,t)}{q_o}}^2-\qty(q_o^2-|q_{gs}(x)|^2) } \,,
	\label{eq:dark_solitons_trap}
	\\
	|q_2(x,t)|^2 = \abs{\sum_n q_b^{(n)}(x,t) }^2 \,.
	\label{eq:bright_solitons_trap}
	\end{gather}
\end{subequations}
In Eq.~(\ref{eq:dark_solitons_trap}), the first term on the right-hand side corresponds to a dark soliton train solution in the absence of a trapping potential having a background amplitude $q_o$, where the product is performed over all the different solutions of a set of zeros $k_o=\{k_{-n},\dots,k_{n}\}$.
The second term properly shapes the former onto the trapped ground state, $q_{\textrm{gs}}(x)$.
Lastly, the absolute value on the right-hand side is introduced so as to ensure the positivity required by the left-hand side.

\begin{figure*}[t]
    \centering
    \includegraphics[width=\linewidth]{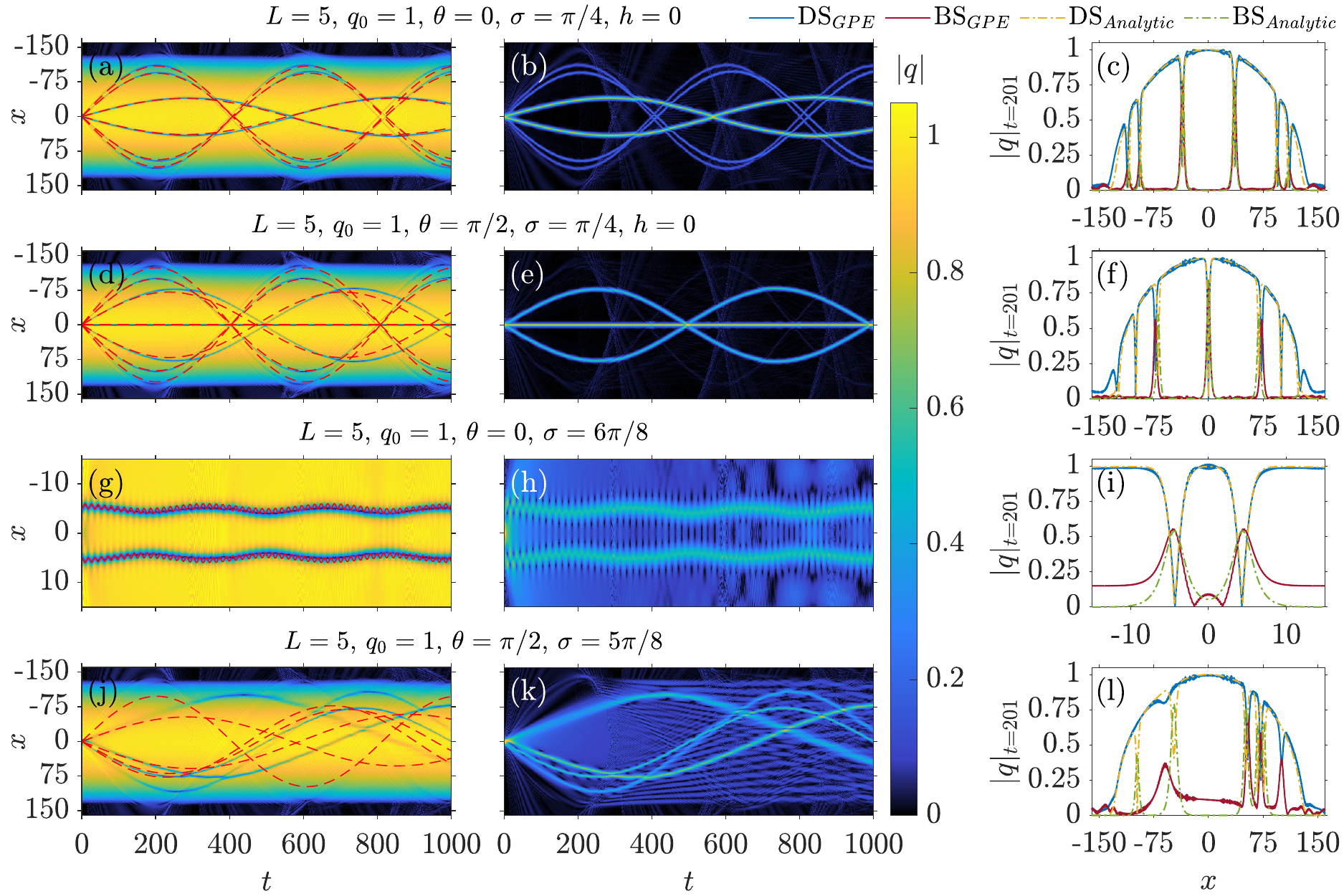}
    \caption{
    Dark-bright solitons generated in the presence of a harmonic trap with a characteristic frequency $\Omega=0.011$ for distinct choices of the involved parameters $L,q_o,\theta,\sigma$ (see legends).
    Each row, from top to bottom, has an initial configuration analogous to the FBTC from Figs.~\ref{fig:GPE_IP}--\ref{fig:GPE_OP_h}, respectively (see Sec.~\ref{sec:homogeneous}).
    Left (middle) column: Spatiotemporal evolution of 
    $|q_1|$ ($|q_2|$) hosting the dark (bright) solitons.
    Red dashed lines correspond to the analytical trajectories [see Eq.~\eqref{eq:trap_trajectory}] using the eigenvalues from the untrapped scenario.
    Right column: Snapshots of $|q_1|$ and $|q_2|$ at $t=201$ given by the CGPE (solid lines) and the analytic in-trap estimates of Eq.~\eqref{eq:dark_solitons_trap} (dashed-dotted lines), for both dark (DS) and bright (BS) soliton counterparts. 
    Note that the quantities shown are measured in transverse oscillator units.
    }
    \label{fig:trap_all}
\end{figure*}

Our results are summarized in Fig.~\ref{fig:trap_all} and Table~\ref{tab:freqs}.
In Fig.~\ref{fig:trap_all} we show the spatiotemporal evolution of $|q_1|$ (left column) and $|q_2|$ (middle column), each of which hosts, respectively, the dark and bright soliton counterparts of the dynamically generated DB solitons.
Additionally, together with $|q_1|$ are depicted the DB soliton trajectories provided by Eq.~\eqref{eq:trap_trajectory} using the eigenvalues of the homogeneous solutions presented in Sec.~\ref{sec:homogeneous} (dashed red lines).
Note here that each row corresponds to a different set of parameters, but with $L=5$ and $q_o=1$ fixed.
For clarity, the dynamical evolution of the DB solitons formed is monitored up to times $t=1000$ but the solitons remain intact while oscillating for times up to $t=3000$.
To offer a head-on comparison between the numerical results and the analytical in-trap estimates of Eq.~\eqref{eq:DB_solitons_trap} we also show a snapshot of $|q_1|$ and $|q_2|$ at $t=201$ (right column) where both the numerical and the analytical results are placed on top of each other.

In Table~\ref{tab:freqs} the analytically obtained oscillation frequency, $\omega_o$, of each DB soliton illustrated in Fig.~\ref{fig:trap_all} is compared with the corresponding numerically identified frequency, $\omega_{\textrm{num}}$.
The latter is measured by following the c.m. of each DB soliton and performing a fast Fourier transform on each obtained trajectory. 
In some cases, however, the presence of radiation hindered tracing the DB soliton c.m. and a manual fitting of $\omega_\textrm{num}$ was required.
Since $\omega_o$ mostly depends on the number of particles hosted in the bright soliton, $N_b$ [see Eq.~\eqref{eq:osc_freq}], we also compare $N_b$ to $N_b^\textrm{num}$.
In order to obtain the number of particles of each bright soliton, $N_b^{\textrm{num}}$, from the numerical solution, a numerical integration with the integration limits properly taken around the bright soliton maxima is carried out [see Eq.~\eqref{eq:Nb}].
Yet, in the full-box case scenarios, the presence of a nonzero
background makes the choice of the integration limits difficult, which
adds a slight  error to our calculation.
Overall, in most of the cases the relative error, $\varepsilon_\omega=\abs{\omega_o-\omega_{\textrm{num}}}/\omega_o$ (idem for $N_b$), is pretty low, suggesting that our analytical solutions, obtained by solving the direct scattering problem in the homogeneous setting, are a good approximation to characterize the solutions in the trapped scenario.
Some exceptions are also discussed below. 

\setlength{\tabcolsep}{3pt}
\renewcommand{\arraystretch}{1.2}
\begin{table*}[t!]
	\centering
	\begin{tabular}{c c c c @{\hspace{1cm}} c c c c @{\hspace{1cm}} c c c c @{\hspace{1cm}} c c c c}
		
		\hline\hline
		\multicolumn{4}{c@{\hspace{1cm}}}{$h=0$, $\sigma=\pi/4$, $\theta=0$} &
		\multicolumn{4}{c@{\hspace{1cm}}}{$h=0$, $\sigma=\pi/4$, $\theta=\pi/2$} &
		\multicolumn{4}{c@{\hspace{1cm}}}{$\sigma=6\pi/8$, $\theta=0$ \footnote{See the discussion around Eqs.~\eqref{eq:beating}} } &
		\multicolumn{4}{c}{$\sigma=5\pi/8$, $\theta=\pi/2$} \\
		\hline\hline
		$k_o$ & $\omega_o$ & $\omega_\textrm{num}$ & $\varepsilon_w$ &
		$k_o$ & $\omega_o$ & $\omega_\textrm{num}$ & $\varepsilon_w$ &
		$k_{\pm1}$ & $\omega_o$ & $\omega_\textrm{num}$ & $\varepsilon_w$ &
		$k_o$ & $\omega_o$ & $\omega_\textrm{num}$ & $\varepsilon_w$ \\
		\hline
		& & & &
		$k_0$ & 0 & 0 & 0 &
		$\omega_{\beta}$ & 0.3539 & 0.3537 & 0.0006 &
		$k_{-2}$ & 6.958 & 6 & 0.14
		\\
		$k_{\pm1}$ & 5.615 & 5.548 & 0.025 &
		$k_{\pm1}$ & 6.670 & 6.35  & 0.048 &
		$\omega_{OP}$ & 0.0188 & 0.0195 & 0.032 &
		$k_{-1}$ & 6.578 & 4.575 & 0.30
		\\
		$k_{\pm2}$ & 7.645 & 7.745 & 0.013 &
		$k_{\pm2}$ & 7.778 & 7.989 & 0.027 &
		& & & &
		$k_0$ & 5.323 & 6.283 & 0.18
		\\
		$k_{\pm3}$ & 7.704 & 7.813 & 0.014 &
		$k_{\pm3}$ & 7.775 & 8.015 & 0.031 &
		& & & &
		$k_1$ & 5.523 & 3.725 & 0.33 
		\\
		\hline
		$k_o$ & $N_b$ & $N_b^{\textrm{num}}$ & $\varepsilon_{N_b}$ &
		$k_o$ & $N_b$ & $N_b^{\textrm{num}}$ & $\varepsilon_{N_b}$ &
		$k_{\pm1}$ & $N_b$ & $N_b^{\textrm{num}}$ & $\varepsilon_{N_b}$ &
		$k_o$ & $N_b$ & $N_b^{\textrm{num}}$ & $\varepsilon_{N_b}$ \\
		\hline
		& & & &
		$k_0$ & 2.636 & 2.648 & 0.005 &
		      & 0.695 & 0.790 & 0.14 &
		$k_{-2}$ & 0.815 & 0.887 & 0.088
		\\
		$k_{\pm1}$ & 2.182 & 2.201 & 0.008 &
		$k_{\pm1}$ & 1.098 & 1.120 & 0.020 &
		& & & &
		$k_{-1}$   & 1.188 & 1.277 & 0.075
		\\
		$k_{\pm2}$ & 0.135 & 0.137 & 0.009 &
		$k_{\pm2}$ & 0.0001 & 0.0003 & 2 &
		& & & &
		$k_{0}$    & 2.511 & 2.524 & 0.005
		\\
		$k_{\pm3}$ & 0.0756 & 0.0831 & 0.099 &
		$k_{\pm3}$ & 0.0035 & 0.0034 & 0.034 &
		& & & &
		$k_{1}$ & 2.283 & 2.601 & 0.13
		\\
		\hline\hline
	\end{tabular}
	\caption{Comparison between the analytically and numerically obtained oscillation frequencies, $\omega_o$ and $\omega_{\textrm{num}}$, and the number of particles of a bright soliton, $N_b$ and $N_b^{\textrm{num}}$, respectively, for each identified DB soliton solution shown in Fig.~\ref{fig:trap_all}.
	From left to right, each column set corresponds, from top to bottom, to each row in Fig.~\ref{fig:trap_all}.
	Each soliton pair $k_{\pm i}$, with $i=1,2,\dots$, is identified using the notation introduced in Sec.~\ref{sec:homogeneous}.
	The relative error is defined as $\varepsilon_\omega=\abs{\omega_o-\omega_{\textrm{num}}}/\omega_o$ (idem for $\varepsilon_{N_b}$).
	The frequencies $\omega_o$ and $\omega_{\textrm{num}}$ have an additional $\times10^3$ factor.
	Other parameters used are $L=5$, $q_o=1$, and $\Omega=0.011$.
	Note that the quantities shown are measured in transverse oscillator units (see text).
	}
	\label{tab:freqs}
\end{table*}

\subsubsection{\textbf{Zero-Box configuration}} \label{sec:zero-box_trap}

The first case example, shown in Figs.~\ref{fig:trap_all}(a)--(c), corresponds to an initial IP ($\theta=0$) zero-box configuration ($h=0$) with $\sigma=\pi/4$, analogous to the homogeneous case shown in Fig.~\ref{fig:GPE_IP}.
Here, three pairs of DB solitons are generated, as expected.
Moreover, the motion of each DB soliton is near perfectly captured by Eq.~\eqref{eq:trap_trajectory}, as depicted by the dashed red lines in Fig.~\ref{fig:trap_all}(a).
Also, in Fig.~\ref{fig:trap_all}(c) we find a very good match between the numeric and analytic DB solitons, bearing our in-trap estimate solution~\eqref{eq:DB_solution_trap}.

The second case corresponds to an initial OP ($\theta=\pi/2$) zero-box configuration ($h=0$) with $\sigma=\pi/4$.
The latter is almost analogous to the homogeneous case shown in Fig.~\ref{fig:GPE_OP}, featuring a static DB soliton formed at the center of the trap, surrounded by a pair of DB solitons and two pairs of (almost) pure dark solitons.
The resulting dynamics are shown in Figs.~\ref{fig:trap_all}(d)--(f).
In Fig.~\ref{fig:trap_all}(d), the analytic trajectories capture pretty well the dynamics of the two most external pairs of \textit{dark} solitons.
Recall that in the homogeneous scenario the fastest DB soliton pair ($k_3$) presented a non-zero bright counterpart, almost nonexistent in Fig.~\ref{fig:trap_all}(e).
Additionally, the in-trap estimates present a very good agreement with the numerical results.
A noticeable discrepancy concerns the central pair of DB solitons.
The comparison between $\omega_o$ and $\omega_o^\textrm{num}$ for this pair is shown in the second column set of Table~\ref{tab:freqs} (see $k_{\pm1}$).
Despite the relative error being not greater than 5$\%$, the long-time dynamics clearly captures its effect.

Also, although we use analytical estimates to describe the in-trap dynamics, a possible source of error is $N_b$ [see Eq.~\eqref{eq:osc_freq}].
However, for the same DB soliton solution ($k_{\pm1}$), in
Table~\ref{tab:freqs} it is shown that the relative error between $N_b$ and $N_b^\textrm{num}$ is of about $2\%$.
The latter suggests that additional sources of error might be present.
For instance, the emitted radiation produced during the interference process might be taken into account.
In this sense, some approximations to Eq.~\eqref{eq:initial_conditions}, e.g., the sigmoid function, have been used to smoothen the step-like shape of the box, decreasing the amount of emitted radiation and showing a small improvement towards the analytical solution (dynamics not shown for brevity).

\subsubsection{\textbf{Full-box configuration}} \label{sec:full-box_trap}

In Sec.~\ref{sec:homogeneous}, we found how a homogeneous setup with an initial full-box configuration, where the two components overlap inside the box, leads to the presence of a nonzero background in the component hosting bright solitons (see Figs.~\ref{fig:GPE_IP_h} and \ref{fig:GPE_OP_h}).

In Figs.~\ref{fig:trap_all}(g)--\ref{fig:trap_all}(i) we present the dynamics resulting from an initial IP ($\theta=0$) full-box configuration with $\sigma=6\pi/8$, which is the in-trap analog of the homogeneous case example shown in Fig.~\ref{fig:GPE_IP_h}.
The homogeneous case resulted into a pair of almost static DB solitons ($v\sim0$) traveling nearly parallel to each other and performing oscillations around their own c.m., i.e., beating.
Here, we identified the same pair of beating DB solitons.
Moreover, their beating behavior can be characterized by the
following expression \cite{Yan2012}
\begin{align}
	\omega_\beta=\frac{1}{2}(\kappa^2 + D^2) \,,
	\label{eq:beating_Yan}
\end{align}
with $\kappa^2=v^2$ and $D^2=\mu\cos^2\phi-\eta^2=A_d^2-A_b^2$.
Using the expressions from Eq.~\eqref{eq:DB_parameters} we can rewrite Eq.~\eqref{eq:beating_Yan} in terms of $z_o$,
\begin{align}
    \omega_\beta = 2(\Re z_o)^2 + \frac{1}{2}(\Im z_o)^2\,,
    \label{eq:beating}
\end{align}
yielding $\omega_\beta=0.3539$.
On the other hand, we numerically followed the CM of our DB soliton pair during the dynamics using the previous procedure described and obtained $\omega_\beta^\textrm{num}=0.3537$.
Comparing $\omega_\beta$ with $\omega_\beta^\textrm{num}$, we find an
extremely good agreement.

Furthermore, in the presence of a harmonic confinement an additional oscillation mode is present in the dynamics, driving both DB solitons to perform out-of-phase oscillations around the center of the trap.
In particular, the out-of-phase mode of the oscillations stems from the presence of the trap and the DB-DB soliton repulsive interaction, characteristic of DB soliton pairs with in-phase bright counterparts~\cite{Segev1998}.
Of course, Eq.~\eqref{eq:trap_trajectory} assumes an oscillation frequency for single DB solitons, and thus it cannot provide a valid description of the motion of this DB soliton pair because it is coupled.

Nonetheless, in Ref.~\cite{Katsimiga2017a} explicit expressions of the energy of the interactions of a pair of DB solitons is provided.
This allows us to derive the expression of the forces involving the dark-dark, bright-bright, and dark-bright interactions, $F_{jk}(x)=-\partial_xE_{jk}(x)$ where $j,k=\{D,B\}$, and numerically solve the equations of motion for our particular DB soliton pair, i.e., $\Ddot{x}=-\omega_o^2x-F_{DD}(x)-F_{BB}(x)-2F_{DB}(x)$.
By doing so, we obtain the trajectory of the DB soliton pair and find the out-of-phase oscillation frequency, $\omega_{OP}=0.0188$, which nicely captures the numerically identified one $\omega_{\textrm{OP}}^{\textrm{num}}=0.0195$.
The latter presents only a relative error $\varepsilon_{OP}=3\%$.
Therefore, we can fully characterize the trajectories of the beating pair of DB solitons by the following expression:
\begin{align}
    x_\pm(t) = &\mp A_{\beta}\cos(\omega_\beta t + \varphi_\beta) \nonumber \\
    &\pm A_\textrm{OP}\cos(\omega_\textrm{OP} t + \varphi_\textrm{OP}) \pm x_o \,,
    \label{eq:trap_beating}
\end{align}
where $A_{\beta,\textrm{OP}}$ denote the amplitude of the beating and out-of-phase oscillations, respectively, and $\varphi_{\beta,\textrm{OP}}$ are additional phases.
Although  the expressions provided in Ref.~\cite{Katsimiga2017a} were derived by means of perturbation theory and predict the oscillation frequency and amplitude of small perturbations, they still provide a good approximation for $\omega_\textrm{OP}$ in this case.
On the contrary, since perturbation theory cannot provide the amplitude of oscillation, we fitted $A_{\beta,\textrm{OP}}$ in Eq.~\eqref{eq:trap_beating} to obtain the trajectories in Fig.~\ref{fig:trap_all}(g).
We also set $\varphi_{\beta,\textrm{OP}}=0$.

It is worth noticing in Fig.~\ref{fig:trap_all}(i), also in this case, the good performance of our analytic in-trap estimates at capturing both the DB soliton profiles, regardless of the presence of the background.

Lastly, we comment on the dynamics of an initial OP ($\theta=\pi/2$) full-box configuration with $\sigma=5\pi/8$.
The resulting spatiotemporal evolutions of $|q_1|$ and $|q_2|$ are shown in Figs.~\ref{fig:trap_all}(j) and \ref{fig:trap_all}(k), respectively, and snapshots of $|q_1|$ and $|q_2|$ at $t=201$ are depicted in Fig.~\ref{fig:trap_all}(l).
First, one can notice that, in Fig.~\ref{fig:trap_all}(j), the analytic solutions (red dashed lines) fail to appropriately capture the dynamics of the DB solitons.
By inspecting once more the analogous homogeneous case shown in Fig.~\ref{fig:GPE_OP_h}, it is observed that the main quantities, i.e., $v$, $A_d$ and $A_b$ [see Figs.~\ref{fig:GPE_OP_h}(e)--\ref{fig:GPE_OP_h}(g), respectively], are still way off from their asymptotic values at $t=250$.
Consequently, the generated DB solitons monitored in the dynamics do not correspond to the analytically expected ones since the former started the in-trap oscillations at earlier times than $t=250$, which interrupted their natural approach to the expected asymptotic solutions.
For instance, from the expected five DB soliton solutions only four are dynamically generated and, as mentioned above, $\omega_o$ and $\omega_\textrm{num}$ differ significantly, with errors well above 14$\%$.

Nevertheless, with an appropriate fit of the parameters to Eq.~\eqref{eq:trap_trajectory}, it can be shown that despite not having the predicted DB solitons, the dynamically formed structures perfectly follow the DB soliton trajectories (fitting not shown for brevity).
Additionally, the fitted parameters applied to our analytical
estimates provide a very
accurate description of the DB soliton profiles.
However, for consistency, in Figs.~\ref{fig:trap_all}(j)--(l) we compare the numerically obtained results with the analytical ones, rather than with the fitted estimates.

For completeness, we also considered in-trap dynamics beyond the Manakov limit, i.e., $g_{jk}\neq1$ (results not shown here for brevity).
In particular, and motivated by relevant studies such as those of Refs.~\cite{Mertes2007,Egorov2013}, we first used for the intracomponent and intercomponent interaction strengths $g_{11}=1.004$, $g_{22}=0.95$, and $g_{12}=g_{21}=0.98$, respectively, corresponding to a system of $^{87}$Rb atoms in the $\ket{1,-1}$ and $\ket{2,1}$ hyperfine states. 
This choice of parameters corresponds to a weakly immiscible mixture, i.e., $g_{11}g_{22}<g_{12}g_{21}$.
Additionally, we also considered a weakly miscible regime, $g_{11}g_{22}>g_{12}g_{21}$, by tuning $g_{12}=0.95$.
Experimentally this could be achieved by means of a Feshbach resonance~\cite{Chin2010}.

In both cases the results are qualitatively similar to the ones presented in the Manakov limit (see Fig.~\ref{fig:trap_all}), and the dark-bright soliton structures emerging in these more realistic setups survive even for long times.
Not only that, but the overall picture is well preserved and the analytical estimates presented in the paper describe with great fidelity most of the cases, at least during the early-time dynamics.
Some of the major differences when comparing these results with the dynamics in the Manakov limit are (i) the presence of a non-negligible amount of noise in the condensates, mostly caused by the overlap of the two-components, and (ii) slightly faster dynamics than those in the Manakov limit.

\section{Conclusions and Future perspectives} \label{sec:conclusions}

In this work, we have investigated the on-demand generation of DB soliton trains arising in a 1D two-component BEC both in the absence and in the presence of a harmonic trap.
We have shown that it is possible to fully characterize a DB soliton array dynamically generated from a box-type initial configuration when a second component is present inside the box.
In particular, we have analytically solved the direct scattering problem for the defocusing VNLS equation utilizing the aforementioned ansatz and obtained expressions for the discrete eigenvalues of the scattering problem.
The latter are directly related to the amplitudes and velocities of the conforming DB solitons and allowed us to construct the exact DB soliton waveforms making use of the IST.

In order to better understand the role of the geometry of the initial box-type configuration in the generation of DB solitons, we explored a wide range of parametric selections.
In general, a wider box generates a higher number of DB soliton structures.
However, the presence of the second component inside the box hinders the appearance of such entities, compared to the single-component case.
If instead both components are present inside the box, the intercomponent interactions practically prevent the emergence of soliton structures unless the presence of the second component overcomes the presence of the first one.
Moreover, we also investigated the effect of a possible phase difference between the distinct regions of the box.
If all regions are in-phase, the number of solitons formed is even, and all of them are paired.
Specifically, each pair consists of DB solitons that share the same characteristics but travel with opposite velocities.
On the contrary, when the sides of the box are out-of-phase, the number of DB solitons is odd and at least one DB soliton appears to be unpaired.
In particular, if the second component is the only one present in the inner box region, the unpaired DB soliton is static.
However, if the majority component is also inside the box, there exists an extra phase-jump at the inter-phase separating the inner and the outer regions of the box, breaking the phase symmetry of the system and leading to the creation of asymmetric DB soliton arrays.
In such a situation, all solutions are unpaired and the number of solitons formed depends on the presence of the components inside the box.

To test our analytical findings we performed direct numerical integration of the multi-component system at hand.
In all the cases in the absence of confinement, we have found that the dynamically produced solitons approach asymptotically the analytically predicted DB amplitudes and velocities. 
In those cases where the initial configuration mixes both components inside the box, we found that the intercomponent interaction stimulates the presence of a finite background surrounding the bright solitons, which leads to the emergence of other exotic structures such as beating DB solitons.
Moreover, we also designed approximate expressions using the analytical solutions of the homogeneous setup to describe the dynamics of DB solitons in the presence of a harmonic trap. Also, we provided expressions for the oscillations of the amplitudes of the dark and bright solitons.
Our estimates showed in most cases a remarkably good agreement with the observed dynamics, with deviations not larger than $5\%$.

An immediate extension of this work points towards richer systems, e.g., spinor BECs~\cite{Kawaguchi2012,Stamper-Kurn2013,Katsimiga2021}.
These systems are already experimentally
realizable~\cite{Stamper-Kurn2001,Chang2004,Chang2005}, and several
works have already exposed the existence of stable solitonic
structures both experimentally~\cite{Bersano2018} and
theoretically~\cite{Nistazakis2008,Xiong2010,Romero-Ros2019,Schmied2020a,Abeya2021}. %
Yet, another possibility for future study is the construction of more complex initial configurations, consisting, for example, of multiple boxes in order to mimic phase structures such as the dark-antidark solitons realized in the experiments of Refs.~\cite{Hamner2013,Katsimiga2020}.
The
latter case, however, requires the scenario of miscibility between the
two components. Finally, the generalization of considerations to
higher
dimensions and, e.g., vortex-bright solitons
therein~\cite{Kevrekidis2016},
could be another fruitful direction for future exploration.

\section*{Acknowledgements}
This work was funded by the Deutsche Forschungsgemeinschaft (DFG, German Research Foundation) – SFB-925 – Project No. 170620586.
This work is supported (G.C.K. and P.S.) by the Cluster of Excellence ``Advanced Imaging of Matter'' of the Deutsche Forschungsgemeinschaft(DFG) - EXC 2056 - project ID 390715994. 
This material is based upon work supported by the US National Science Foundation under Grants No. PHY-2110030
and No. DMS-1809074 (P.G.K.), as well as No. DMS-2009487 (G.B.) and No. DMS-2106488 (B.P.).

\appendix
\section{Further insights of the DB soliton solutions}\label{app:DB_solutions}

Since we are finding the eigenvalues as zeros of $s_{11}(k)$, it is important to relate $\Re k_o$ and $\Im k_o$ to $z_o$.
From the definition of the uniformization variable one has $z_o=k_o+\lambda(k_o)$, but this relationship requires dealing with the branches of $\lambda(k_0)$.
However, this can be bypassed as follows.
From Eqs.~\eqref{eq:inverted_z} we have
\begin{subequations}
	\label{eq:Re_Im_k}
	\begin{align}
	\Re k =\frac{1}{2}\left(1+\frac{q_o^2}{|z|^2} \right)\Re z  \,, 
	\\
	\Im k =\frac{1}{2}\left(1-\frac{q_o^2}{|z|^2} \right)\Im z \,,
	\end{align}
\end{subequations}
and
\begin{subequations}
	\label{eq:Re_Im_lambda}
	\begin{align}
	\Re \lambda =\frac{1}{2}\left(1-\frac{q_o^2}{|z|^2} \right)\Re z \,,
	\\
	\Im \lambda =\frac{1}{2}\left(1+\frac{q_o^2}{|z|^2} \right)\Im z \,.
	\end{align}
\end{subequations}
The second relation shows that $\Im \lambda>0 \iff \Im z>0$, which restricts the eigenvalues as zeros of $s_{11}(z)$ in the upper-half plane of $z$, and $\beta \in (0,\pi]$.
Additionally, when $\Im z>0$, $\abs{z}<q_o \iff \Im k < 0$.
Thus, given that the upper half of the circle of radius $q_o$ in the $z$-plane is in one-to-one correspondence with the lower half plane of the upper sheet of the Riemann surface, $k_o$ eigenvalues can have any $\Re k$ and $\Im k<0$, provided that $\Im \lambda(k) >0$.
Note that the latter differs from the scalar case  of Ref.~\cite{Romero-Ros2021} where $-q_o<k<q_o$.

In Eq.~\eqref{eq:DB_parameters_gamma} it remains to express $\gamma$ in terms of $k_o$, which can be done as follows.
Let us for brevity introduce $x=\Re k_o$ and $y=\Im k_o$.
Then, from Eqs.~\eqref{eq:Re_Im_k} one has $\Re z_o=2x/(1+\gamma^2)$, $\Im z_o=2y/(1-\gamma^2)$, and $|z_o|^2=q_o^2/\gamma^2=(\Re z_o)^2+(\Im z_o)^2$ which upon substitution yields
\begin{align}
4\frac{x^2}{(1+\gamma^2)^2}+4\frac{y^2}{(1-\gamma^2)^2}=\frac{q_o^2}{\gamma^2} \,,
\end{align}
namely a (simplified) quartic equation for $\Gamma\equiv\gamma^2$
\begin{widetext}
	\begin{align}
	\Gamma^4
	-\frac{4}{q_o^2}(x^2+y^2)\Gamma^3 -2(1-\frac{4}{q_o^2}x^2+\frac{4}{q_o^2}y^2)\Gamma^2 
	-\frac{4}{q_o^2}(x^2+y^2)\Gamma+1=0  \,.
	\label{eq:Gamma_quartic}
	\end{align}
	The solutions of Eq.~\eqref{eq:Gamma_quartic} are:
	\begin{subequations}
		\label{eq:gamma_solutions}
		\begin{align}
		q_o^2 \gamma_{\pm}^2=|k_o|^2-\beta \pm\sqrt{2}\sqrt{|k_o|^4-2q_o^2(\Re k_o)^2+|k_o|^2(q_o^2-\beta) } \,,  \label{eq:gamma_solutions_complex}
		\\
		q_o^2 \gamma_{\pm}^2=|k_o|^2+\beta \pm\sqrt{2}\sqrt{|k_o|^4-2q_o^2(\Re k_o)^2+|k_o|^2(q_o^2+\beta) } \,, \label{eq:gamma_solutions_real}
		\end{align}
	\end{subequations}
\end{widetext}
with
\begin{align}
\beta=\sqrt{(q_o^2+|k_o|^2)^2-4q_o^2 \Re^2 k_o} \,.
\end{align}
The pair of solutions in  Eq.~\eqref{eq:gamma_solutions_complex} are complex conjugate, while those in Eq.~\eqref{eq:gamma_solutions_real} are real.
We are interested in real solutions with $\gamma>1$, which are then given by $\gamma_+$ in Eq.~\eqref{eq:gamma_solutions_real}.
Notice that $\gamma_+$ involves only real square roots, thus avoiding complex branches.
Hence, using $\gamma_+$ in Eq.~\eqref{eq:DB_parameters_gamma} provides all the soliton parameters in terms of $k_o=\Re k_o+i\Im k_o$ for arbitrary $\Re k_o \in \mathbb{R} $ and $\Im k_o<0$.

At this point, it is also possible to retrieve the soliton parameters for the single-component case.
Recall that, for the scalar defocusing NLS equation, the zeros are real and simple, belonging to the spectral gap $k\in(-q_o,q_o)$~\cite{Faddeev2007,Romero-Ros2021}.
This {directly implies} that $|z_o| = q_o \:\forall\: k_o$.
Therefore, Eqs.~\eqref{eq:DB_parameters} read
\begin{subequations}
	\label{eq:DS_parameters} 
	\begin{align}
	A_d &= q_o\sin\beta_o \equiv \sqrt{q_o^2-k_o^2} \,, 
	\\
	A_b &= 0 \,, 
	\\
	v &= -2q_o\cos\beta_o \equiv -2k_o \,.
	\end{align} 
\end{subequations}

For completeness, we note here that it is also possible to obtain the zeros $k_o$ given the soliton parameters $A_d,A_b$ and $v$.
In particular, using Eqs.~\eqref{eq:DB_parameters} we obtain
\begin{subequations}
	\begin{gather}
	A_b^2 = A_d^2\qty(1-\frac{\abs{z_o}^2}{q_o^2}) \,,
	\\
	\cos\beta_o = \pm\sqrt{1-\frac{A_d^2}{q_o^2}} \,,
	\\
	\sin\beta_o = \frac{A_d}{q_o} \,.
	\end{gather}
\end{subequations}
Recalling now that
\begin{align}
z_{o \pm} & \equiv \abs{z_o}(\cos\beta_o + i\sin\beta_o) \nonumber
\\
& = q_o\sqrt{1-\frac{A_b^2}{A_d^2}}\qty(\pm\sqrt{1-\frac{A_d^2}{q_o^2}} + i\frac{A_d}{q_o}) \,,
\label{eq:z_param}
\end{align}
$\abs{z_o}<q_o$ is automatically satisfied and the sign of $\cos\beta_o$ is determined by Eq.~\eqref{eq:v}.
If $v>0$ then $\cos\beta_o > 0$, while if $v<0$ then $\cos\beta_o < 0$.
Now, substituting Eq.~\eqref{eq:z_param} into Eq.~\eqref{eq:Re_Im_k} yields
\begin{subequations}
	\begin{align}
	\Re k_o &= \text{sgn}(v)\frac{q_o}{2}\sqrt{1-\frac{A_d^2}{q_o^2}}\times \nonumber \\
	&\hspace{23.5pt} \times \qty[\qty(1-\frac{A_b^2}{A_d^2})^{-\frac{1}{2}} + \qty(1-\frac{A_b^2}{A_d^2})^{\frac{1}{2}}] \,,
	\\
	\Im k_o &= -\frac{A_d}{2}\qty[\qty(1-\frac{A_b^2}{A_d^2})^{-\frac{1}{2}} - \qty(1-\frac{A_b^2}{A_d^2})^{\frac{1}{2}}] \,.
	\end{align}
\end{subequations}
It is clear from the above expression that $\Im k_o < 0$, and since $\Im z_o >0$ it follows that $\Im \lambda (k_o) >0$.

\section{Dark-bright soliton amplitudes in the presence of a harmonic trapping potential}\label{app:amplitudes}

\begin{figure}[t]
    \centering
    \includegraphics[width=\columnwidth]{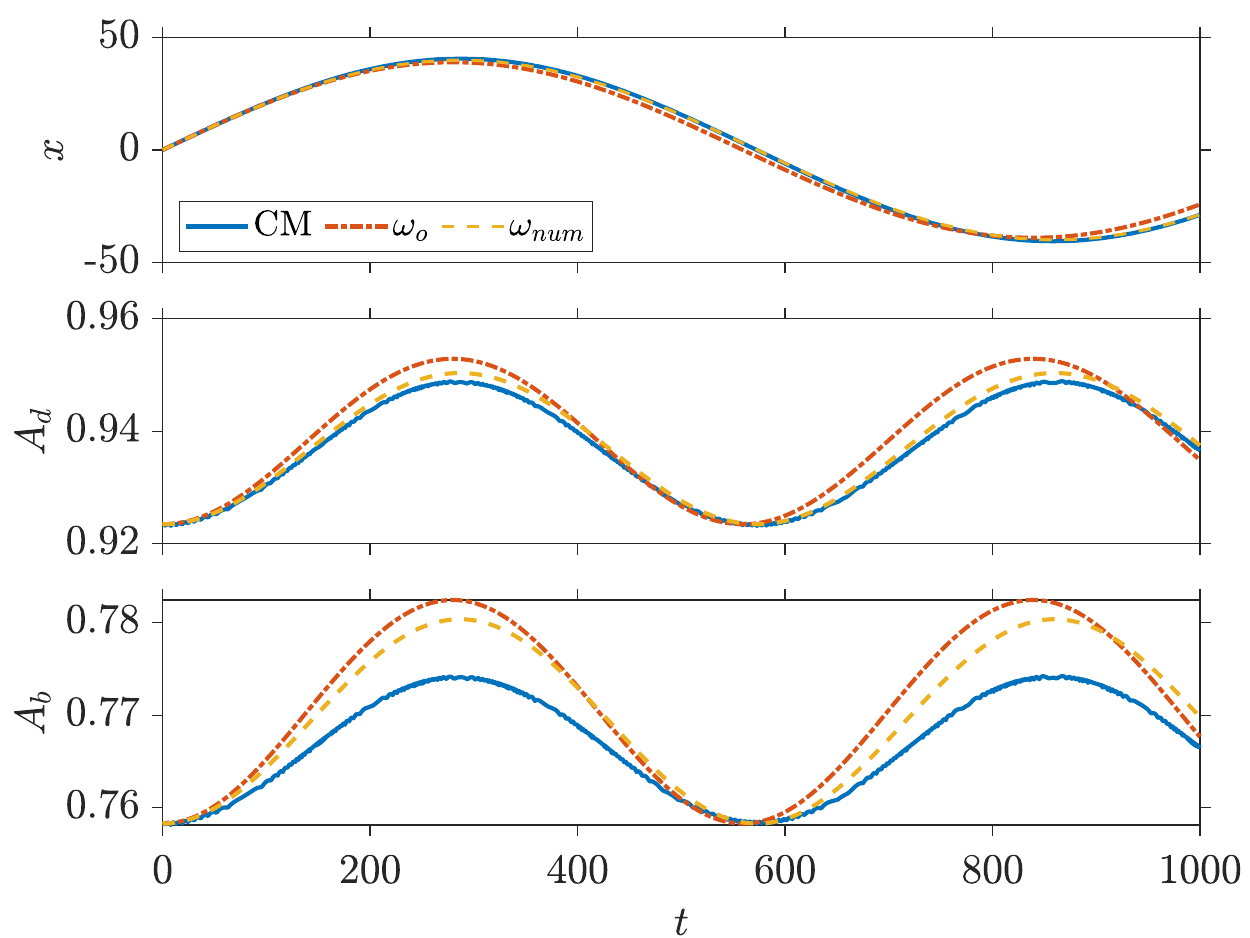}
    \caption{
    Trajectory, $x$, and dark, $A_d$, and bright, $A_b$, amplitudes of the DB soliton solution $k_1$ shown in Fig.~\ref{fig:GPE_IP}.
    The numerical magnitudes, obtained by following the c.m. (solid blue line), are compared to the analytical estimates in Eq.~\eqref{eq:trap_trajectory} and in Eqs.~\eqref{eq:DB_parameters} [with $\beta_o\rightarrow\beta(t)$] given by the analytical in-trap oscillation frequency, $\omega_o$ [see Eq.~\eqref{eq:osc_freq}] (dashed-dotted red lines), and the numerically obtained one, $\omega_\textrm{num}$ (dashed yellow lines).
    Note that the quantities shown are measured in transverse oscillator units.
    }
    \label{fig:k1_IP}
\end{figure}

One important characteristic of solitons is that they preserve their shape.
Also, it is well known that, in the presence of a harmonic trapping potential, DB solitons can undergo oscillations of frequency $\omega_o$ [see Eq.~\ref{eq:osc_freq}].
However, here we found that DB solitons change size as they perform such oscillations in the trap.
This particular feature is attributed to the intercomponent interaction, $g_{12}$, coupling the dark and bright counterparts, and to their constraints with the DB soliton velocity.
Below we derive the expressions to describe such amplitude oscillations, but the role of $g_{12}=1$ will be hidden in the equations.

At the turning points of their oscillatory trajectories ($x_t=\pm \Re z_o/\omega_o$) the DB soliton velocity must be 0, which implies that its amplitudes are maximal [see Eqs.~\eqref{eq:DB_parameters}].
In particular, for the dark counterpart that resides on top of the density background of the condensate $A_{d\textrm{(max)}}^2=|q_{gs}(x_t)|^2$.
Following the same lines, at the center of the trap $(x=0)$ the velocity of the DB soliton is maximal, and thus its amplitudes are minimal and, more precisely, coincide with those of the homogeneous setup, i.e., $A_{d\textrm{(min)}}^2=q_o^2\sin^2\beta_o$. 

Having now at hand the extremes of $A_d$, only the frequency of such oscillations is missing.
In this case, it is enough to notice that in half of a trap oscillation period the dark amplitude would perform a full cycle.
Therefore, it is straightforward to express the amplitude of the dark counterpart as
\begin{align}
    A_d^2(t) = &\frac{1}{2}(A_{d\textrm{(max)}}^2+A_{d\textrm{(min)}}^2) \nonumber \\
    - &\frac{1}{2}(A_{d\textrm{(max)}}^2-A_{d\textrm{(min)}}^2)\cos{(2\omega_o t)} \,,
\end{align}
which after some algebra yields
\begin{align}
    A_d^2(t) = q_o^2\sin^2{\beta_o}\cos^2(\omega_o t) + |q_{gs}(x_t)|^2\sin^2(\omega_o t) \,.
    \label{eq:Ad(t)}
\end{align}

Now, comparing Eq.~\eqref{eq:Ad} to Eq.~\eqref{eq:Ad(t)}, we obtain 
\begin{align}
    \sin^2\beta(t)=\sin^2{\beta_o}\cos^2(\omega_o t) + \frac{|q_{gs}(x_t)|^2}{q_o^2}\sin^2(\omega_o t) \,,
    \label{eq:beta_t_app}
\end{align}
which is equivalent to Eq.~\eqref{eq:beta_t}, as shown below.
From here, by replacing $\beta_o\rightarrow\beta(t)$ in Eqs.~\eqref{eq:DB_solution}, \eqref{eq:xi_nu}, and \eqref{eq:DB_parameters}, the DB soliton solution for in-trap oscillations follows [see Eq.~\eqref{eq:DB_solution_trap}].

It is also important to notice that $|z(t)|=|z_o|$ does not change over time, since the uniformization parameter $z$ is unique to each DB soliton.
Additionally, $\beta(t)$ satisfies the condition required by $0<\beta_o\leq\pi$ which restricts the eigenvalues in the upper-half plane of $z$.
For instance, $\beta(t)=\arcsin{\frac{A_d(t)}{q_o}}$ and, since $0<A_d(t)\leq q_o \;\forall\; t$, then $0<\beta(t)\leq\pi/2 \;\forall\; t$.
Note that the values $\pi/2<\beta(t)\leq\pi$, which are missing due to the $\arcsin(\dots)$, only affect the sign of the velocity of the soliton~\eqref{eq:v}.
However, Eq.~\eqref{eq:v} is not valid to define the DB soliton velocity in the presence of a trap, which instead is derived from Eq.~\eqref{eq:trap_trajectory}.

One could also try to derive $A_d(t)$ from the velocity of the in-trap oscillations of the DB soliton provided by Eq.~\eqref{eq:trap_trajectory}.
It reads as
\begin{align}
    v(t)\equiv \frac{\dd x_c}{\dd t} = v_o\cos(\omega_o t) \,.
    \label{eq:v(t)}
\end{align}
Then, by comparing Eq.~\eqref{eq:v(t)} to \eqref{eq:v} we obtain
\begin{align}
    \cos\beta(t) = \cos\beta_o\cos(\omega_o t) \,,
    \label{eq:cos_beta(t)}
\end{align}
and, therefore,
\begin{align}
    A_d^2(t) = q_o^2\sin^2\beta(t) = q_o^2-q_o^2\cos^2\beta_o\cos^2(\omega_o t)\,.
    \label{eq:Ad(t)_wrong}
\end{align}
In this case, we see that $A_{d\textrm{(min)}}^2 \leq A_d^2(t) \leq q_o^2$, with $A_{d\textrm{(min)}}^2=q_o^2\sin\beta_o$.
Obviously, $A_d^2(t)$ cannot be equal to $q_o^2$ since $|q_{gs}(x)|^2 \leq q_o^2$ and the only case with $A_d^2(t) = q_o^2$ corresponds to a static dark soliton centered at $x=0$. 
Consequently, deriving $A_d(t)$ from Eq.~\eqref{eq:trap_trajectory} is clearly missing information about the trap geometry.

In particular, it would be enough to add the term $-V(x_t)\sin^2(\omega_o t)$ into Eq.~\eqref{eq:Ad(t)_wrong}, where $x_t=\pm v_o/\omega_o$ is the turning point of the in-trap oscillations of the DB soliton.
After some trivial calculations we recover Eq.~\eqref{eq:Ad(t)}:
\begin{align}
    A_d^2(t) &= q_o^2\sin^2\beta_o\cos^2(\omega_o t) \nonumber \\
             &+ (q_o^2-V(x_t))\sin^2(\omega_o t) \,,
\end{align}
where $(q_o^2-V(x_t))=|q_\textrm{gs}(x_t)|^2$ is the well-known Thomas-Fermi approximation~\cite{thomas1927,fermi1927}.

To adequately approach this problem, we can define a complex trajectory
\begin{align}
    \Tilde{x}(t) = \frac{v_o}{\omega_o}\qty(\sin(\omega_o t) + \frac{i}{\sqrt{2}\gamma}\frac{\Omega}{\omega_o}\cos(\omega_o t))\,,
\end{align}
where the soliton trajectory is $x_c(t) = \Re \Tilde{x}(t)$, and the trap geometry is taken into account by the imaginary term.
From here, we derive $\Tilde{x}(t)$ over time to obtain the (complex) velocity,
\begin{align}
    \Tilde{v}(t) = v_o\qty(\cos(\omega_o t) - \frac{i}{\sqrt{2}\gamma}\frac{\Omega}{\omega_o}\sin(\omega_o t))\,.
    \label{eq:v_tilde}
\end{align}
Then, comparing \eqref{eq:v_tilde} to Eq.~\eqref{eq:v} we obtain our final expression \eqref{eq:beta_t},
\begin{align}
    \cos^2\beta(t) = \cos^2\beta_o\cos^2(\omega_o t)+\frac{1}{2q_o^2}\Omega^2\qty(\frac{v_o}{\omega_o})^2\sin^2(\omega_o t) \,,
    \label{eq:cos2_beta(t)}
\end{align}
containing the information of the trap geometry.
Again, Eq.~\eqref{eq:beta_t_app} can be retrieved by performing an appropriate manipulation of Eq.~\eqref{eq:cos2_beta(t)}.

In order to compare the analytical estimate of Eq.~\eqref{eq:cos2_beta(t)} with numerical DB soliton dynamics, the DB soliton $k_1$ from Fig.~\ref{fig:GPE_IP} is placed alone at the center of a BEC trapped in the harmonic confinement used in this work (see Sec.~\ref{sec:trap}).
Since $v_{k_1}(t=0)\neq0$ it undergoes oscillations.
By following its c.m., we monitor its position, $x$, and its dark, $A_d$, and bright, $A_b$, amplitudes over time.

In Fig.~\ref{fig:k1_IP}, the trajectory and amplitudes of $k_1$ obtained from following its CM (solid blue lines) are compared to the analytical estimates in Eq.~\eqref{eq:trap_trajectory} and in Eqs.~\eqref{eq:DB_parameters} (with $\beta_o\rightarrow\beta(t)$) given by the analytical in-trap oscillation frequency, $\omega_o$ [see Eq.~\ref{eq:osc_freq}] (dashed-dotted red lines), and the numerically obtained one, $\omega_\textrm{num}$ (dashed yellow lines).
Here, the oscillations of $A_d$ and $A_b$ are clearly identified.
Also, our analytical estimates are in good agreement with the numerical findings, with relative errors not larger than $1\%$ at the instant of maximum discrepancy.
In this case we define the relative error as $\varepsilon(A)=|A_{CM}-A_{\omega_\textrm{num}}|/A_{CM}$, which yields $\varepsilon(A_d)=0.0015$ and $\varepsilon(A_b)=0.0079$.

\bibliographystyle{apsrev4-1}
\bibliography{Prinari_Roots.bib}

\end{document}